\documentclass[fleqn,usenatbib]{mnras}

\DeclareRobustCommand{\VAN}[3]{#2}
\let\VANthebibliography\thebibliography
\def\thebibliography{\DeclareRobustCommand{\VAN}[3]{##3}\VANthebibliography}

\usepackage{graphicx}	
\usepackage{amsmath}	
\usepackage{amssymb}	
\usepackage{pdflscape}
\usepackage{multirow}
\usepackage{booktabs}
\DeclareRobustCommand{\ZAK}[3]{#3} 

\usepackage{newtxtext,newtxmath}

\usepackage[T1]{fontenc}

\title[Survey of sodium in exoplanet atmospheres]{A survey of sodium absorption in ten giant exoplanets with high-resolution transmission spectroscopy}

\author[A. Langeveld et al.]{
Adam B. Langeveld,$^{1}$\thanks{E-mail: adam.langeveld@ast.cam.ac.uk}
Nikku Madhusudhan,$^{1}$\thanks{E-mail: nmadhu@ast.cam.ac.uk}
Samuel H. C. Cabot$^{2}$, 
\\
$^{1}$Institute of Astronomy, University of Cambridge, Madingley Road, Cambridge, CB3 0HA, UK\\
$^{2}$Yale University, 52 Hillhouse Avenue, New Haven, CT 06511, USA\\
}

\date{Accepted 2022 May 29. Received 2022 May 2; in original form 2021 October 12}

\pubyear{2022}

\begin{document}
\label{firstpage}
\pagerange{\pageref{firstpage}--\pageref{lastpage}}
\maketitle

\begin{abstract}
The alkali metal sodium (Na) is one of the most commonly detected chemical species in the upper atmospheres of giant exoplanets. In this work we conducted a homogeneous survey of Na in a diverse sample of ten highly irradiated giant exoplanets using high-resolution transmission spectroscopy. Our sample includes nine planets with previous Na detections and one new detection. We confirm previous detections and assess multiple approaches for deriving Na line properties from high-resolution transmission spectra. The homogeneously measured sodium line depths were used to constrain the atmospheric heights ($H_{\text{Na}}$) with respect to the planetary radii ($R_{\text{p}}$). We assess an empirical trend describing the relative atmospheric height ($H_{\text{Na}}/R_{\text{p}}$) as a function of planetary equilibrium temperature ($T_{\text{eq}}$) and surface gravity ($g$), in which $H_{\text{Na}}/R_{\text{p}}$ decreases exponentially with $\xi \propto gT_{\text{eq}}$, approaching a constant at large $\xi$. We also report the sodium D2/D1 line ratios across our sample and find that seven targets have line ratios that are consistent with unity. Finally, we measured net blueshifted offsets of the sodium absorption lines from their rest frame wavelengths for all ten planets, corresponding to day-night wind velocities of a few km~s$^{-1}$. This suggests that the broad sample of exoplanets share common underlying processes that govern atmospheric dynamics. Our study highlights a promising avenue for using high-resolution transmission spectroscopy to further our understanding of how atmospheric characteristics vary over a diverse sample of exoplanets.
\end{abstract}

\begin{keywords}
Planets and satellites: atmospheres -- Planets and satellites: gaseous planets -- Atmospheric effects -- Techniques: spectroscopic -- Methods: observational
\end{keywords}


\section{Introduction}
\label{sec:introduction}

The number of known exoplanets is rapidly increasing, followed by the characterisation of their bulk properties and atmospheres. Many planets are now being extensively studied using a variety of methods, and numerous detections of chemical species in their atmospheres have been reported. With these developments the field is moving towards attempting to answer population-level questions on exoplanetary atmospheres. For example, are there any trends linking macroscopic planetary properties to chemical compositions, and how do atmospheric processes vary over a wide range of planetary properties? 

Spectroscopy of transiting exoplanets has so far proved to be the most effective method for atmospheric characterisation. The geometry of these systems allows for three opportunities to examine the planetary atmospheres: (1) when the planet passes in front of its host star during the primary eclipse, yielding a transmission spectrum \citep{charbonneau2002, redfield2008, snellen2008}; (2) when the planet passes behind the star during the secondary eclipse, yielding an emission spectrum \citep{charbonneau2008, knutson2009}; and (3) when the planet orbits between the primary and secondary eclipse, allowing for acquisition of a phase curve \citep{stevenson2014, demory2016}. For the primary and secondary eclipses, the atmospheric spectrum can be computed by comparing the change in flux observed during and before/after the eclipse \citep{seager2000} -- any differences are likely to be caused by absorption or emission due to chemical species within the atmosphere.

In particular, transmission spectroscopy has been successful at identifying numerous chemical species in a diverse range of exoplanets, through a combination of low-resolution and high-resolution observations at UV, optical, and infrared wavelengths \citep{madhusudhan2019}. Among the most commonly predicted and detected species in gas giant planets are the alkali metals Na and K, both of which have strong optical absorption features \citep[e.g.][]{seager2000, brown2001, charbonneau2002, redfield2008, nikolov2016, sedaghati2016, sing2016, casasayas-barris2017, casasayas-barris2018, wyttenbach2017, chen2018, chen2020b, jensen2018, deibert2019, hoeijmakers2019, seidel2019, cabot2020}. Na absorption in particular is characterised by the strong Na~\textsc{i} doublet lines at 5889.951 and 5895.924~{\AA} \citep{seager2000, burrows2003}.

The first successful detections of sodium with high-resolution transmission spectroscopy were made by \citet{redfield2008} for HD~189733~b and \citet{snellen2008} for HD~209458~b, paving the way towards the development of new techniques suited for analysing data acquired from ground-based instruments. Later, pioneering work led by \citet{wyttenbach2015} demonstrated how telluric contamination and planetary radial velocity shifts can impact atmospheric detections for HD~189733~b, leading to key results such as measurements of the depths of the sodium doublet lines, a strongly increasing temperature gradient, and high-altitude atmospheric winds. This opened up a new avenue for characterising exoplanet atmospheres in the optical domain using 4~m class telescopes and instruments built specifically for exoplanet spectroscopy. Using data acquired with the same instruments, subsequent studies applied the same techniques to analyse sodium absorption in other exoplanets \citep[e.g.][]{casasayas-barris2017, wyttenbach2017, seidel2019, chen2020b}, and additionally detect multiple chemical species using the cross-correlation method \citep[e.g.][]{hoeijmakers2018, hoeijmakers2019, hoeijmakers2020, casasayas-barris2019, yan2019, ben-yami2020, cabot2020, kesseli2021} which was previously successful for infrared observations \citep[][]{snellen2010, brogi2016}.
A number of studies have also sought to understand the theoretical interpretation of observed Na lines \citep[e.g.][]{fortney2003, vidal-madjar2011, heng2015a, gebek2020}.

Motivated by these key results and the wealth of available data, we aim to look for trends linking the properties of the exoplanetary systems to the absorption of chemical species within their atmospheres. In this work, we conduct a homogeneous survey of sodium absorption in a broad sample of ten transiting gas giants which have been observed with the High Accuracy Radial velocity Planet Searcher spectrographs in the southern (HARPS) and northern (HARPS-N) hemispheres. Table~\ref{tab:planet_sample} shows a list of chosen targets, together with the V-band apparent magnitudes of the host stars, the planetary equilibrium temperatures, and the stellar effective temperatures. Equilibrium temperatures were calculated using 
\begin{equation}
    T_{\text{eq}} = T_{\text{eff}}\left[\frac{R_\ast^2}{2a^2}(1-f_\text{r})(1-A_\text{B})\right]^{\frac{1}{4}} ~, 
    \label{eqn:equilibrium_temperature}
\end{equation}
assuming uniform heat redistribution ($f_\text{r} = 0.5$), zero bond albedo ($A_\text{B}=0$), and a circular orbit. All stellar, planetary, and system parameters used within this work can be found in Tables~\ref{tab:parameters1-5} and \ref{tab:parameters6-10} in Appendix~\ref{app:system_parameters}. 

\begin{figure}
    \centering
    \includegraphics[width=\columnwidth]{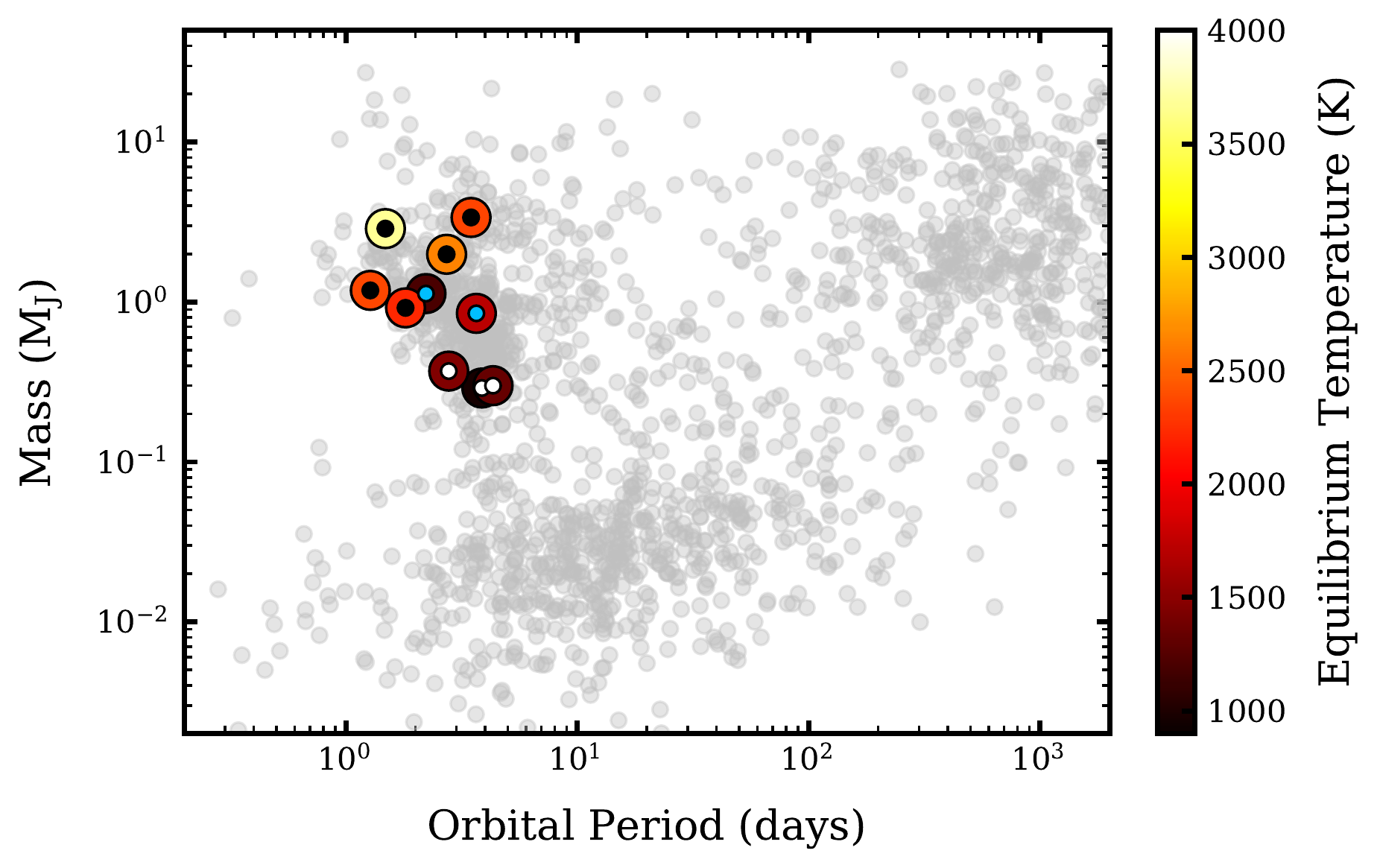}
    \vspace{-1em}
    \caption{Masses and orbital periods of known exoplanets (grey dots). The ten exoplanets analysed in this survey are colour-coded from dark-red to white according to their equilibrium temperature. The sample contains a mix of ultra-hot Jupiters (black dots), hot Jupiters (light-blue dots), and hot Saturns (white dots). Data obtained from the NASA Exoplanet Archive \citep{akeson2013}.}
    \label{fig:mass-period}
\end{figure}

Our sample consists of ten giant exoplanets with diverse properties: three hot Saturns (WASP-69b, WASP-21b, WASP-49b), two hot Jupiters (HD~189733~b, WASP-79b), and five ultra-hot Jupiters (WASP-76b, MASCARA-2b/KELT-20b, WASP-121b, WASP-189b, KELT-9b) -- all of which are tidally locked and strongly irradiated due to the proximity to their host stars. Figure~\ref{fig:mass-period} shows the masses and periods of these planets in comparison to the known exoplanet population.

Ultra-hot Jupiters (UHJs) have equilibrium temperatures greater than $\sim2200$~K \citep{parmentier2018} and even hotter day-sides \citep[e.g.][]{kreidberg2018, yan2018, helling2019a, helling2019b}, making them particularly interesting targets for investigating comparisons between the hottest planets and the coldest stars. Molecules within the hottest day-side regions of the planetary atmospheres can become thermally dissociated into their constituent atoms \citep{arcangeli2018, bell2018, komacek2018, lothringer2018, parmentier2018}, and neutral atomic species can condensate as they circulate to the colder night-sides \citep[e.g. \mbox{WASP-76b:}][]{ehrenreich2020, kesseli2021, wardenier2021}. Hot Jupiters and hot Saturns are similar in size to UHJs, but have cooler equilibrium temperatures ($\lesssim2200$~K) which can lead to different atmospheric chemistry \citep{madhusudhan2012, moses2013, lothringer2018}. When looking for trends, it is important to choose a sample of planets that accurately represents the diversity of these strongly irradiated gas giants to give a clear view of how atmospheric chemistry may change over a broad range of planetary properties.

\begingroup
\renewcommand{\arraystretch}{1.0} 
\begin{table}
    \centering
    \caption{The ten gas giant planets analysed in this survey, with their respective equilibrium temperatures ($T_{\text{eq}}$), stellar effective temperatures ($T_{\text{eff}}$), and stellar V-band apparent magnitudes ($m_{\text{V}}$). Equilibrium temperatures were derived using equation~\ref{eqn:equilibrium_temperature}, 
    assuming uniform heat redistribution ($f_\text{r} = 0.5$), zero bond albedo, and a circular orbit. 
    The system parameters and corresponding references can be found in Tables~\ref{tab:parameters1-5} and \ref{tab:parameters6-10} in Appendix~\ref{app:system_parameters}, and are discussed in section~\ref{sec:observations}. 
    $^\dagger$~MASCARA-2b is also referred to as KELT-20b.}
    \label{tab:planet_sample}
    \begin{tabular*}{\columnwidth}{l@{\extracolsep{\fill}}lll}
    \toprule
    Planet                  & $T_{\text{eq}}$ (K)   & $T_{\text{eff}}$ (K)  & $m_{\text{V}}$ \\
    \midrule                
    WASP-69b                & $960\pm20$            & $4715\pm50$           & 9.873     \\
    HD~189733~b             & $1200\pm15$           & $5052\pm16$           & 7.67      \\
    WASP-21b                & $1330\pm40$           & $5800\pm100$          & 11.59     \\
    WASP-49b                & $1450\pm40$           & $5600\pm150$          & 11.352    \\
    WASP-79b                & $1720\pm30$           & $6600\pm100$          & 10.044    \\
    WASP-76b                & $2210\pm30$           & $6329\pm65$           & 9.518     \\
    MASCARA-2b~$^\dagger$   & $2350\pm50$           & $8980\pm130$          & 7.59      \\
    WASP-121b               & $2360\pm60$           & $6459\pm140$          & 10.514    \\
    WASP-189b               & $2640\pm40$           & $7996\pm99$           & 6.62      \\
    KELT-9b                 & $3670\pm150$          & $9600\pm400$          & 7.55      \\
    \bottomrule
    \end{tabular*}
\end{table}
\endgroup

Previous work has highlighted the importance of correcting for several effects that can impact the quality of the extracted planetary transmission spectrum. First, spectra acquired with ground-based instruments are contaminated by absorption from Earth's atmosphere -- in the optical domain, telluric water and oxygen are the dominant sources of contamination \citep{smette2015, kausch2015, wyttenbach2015, allart2017, langeveld2021}. Additionally, sodium absorption from interstellar sources \citep{casasayas-barris2018, chen2020b, cabot2021} or sodium emission from the sky \citep{casasayas-barris2017} can be detrimental when analysing sodium in exoplanet atmospheres. Next, the spectra must be Doppler shifted to account for stellar, systemic, and planetary radial velocities to make sure that atmospheric features are recovered in the planetary rest frame. The stellar spectral lines may also be distorted as a result of Centre-to-Limb Variation (CLV) \citep{czesla2015, yan2017} and the Rossiter-McLaughlin (RM) effect \citep{rossiter1924, mclaughlin1924, queloz2000, triaud2018}; a planet passing in front of a rotating star blocks out different amounts of blueshifted and redshifted light as it transits across the stellar disc, which can imprint spurious signals within the transmission spectrum. Correcting for these effects will prevent false identifications of chemical species, and is especially critical for future work when newly developed high-resolution spectrographs are used to observe smaller and fainter targets.

The capabilities of high-resolution spectrographs such as HARPS have made it possible to observe planets around faint host stars with V-band apparent magnitudes greater than 11 \citep{wyttenbach2017, chen2020b}. 
However, spectra from these faint targets have a lower signal-to-noise ratio (SNR), particularly inside deep stellar lines (e.g. the Na doublet) where the flux is often $<100$ counts per exposure at the very depth of the line cores.
This drastically increases the difficulty in measuring a change in flux due to the minuscule amount of light absorbed by the planet's atmosphere. Since the transit occurs during a fixed time period, the exposure time cannot simply be increased to enhance the SNR without sacrificing the total number of frames. When the in-transit spectra are divided by the combined out-of-transit spectrum, there will be a band of low-SNR residuals which can mask out absorption features from the planetary atmosphere \citep{seidel2020b}. Further complications arise if telluric sodium emission is present in the same location as the deep stellar line cores, and can result in falsely identified features in the transmission spectrum \citep{seidel2020c}. 
In recent work, assigning weights to the spectra before combining \citep[e.g.][]{allart2020, chen2020b, borsa2021, sedaghati2021} or ignoring data in regions where the planetary signal overlaps with the location of the stellar line cores \citep{seidel2020b, seidel2020c} has proved to be effective at nullifying these effects.

In this work, a standard method is applied across all datasets: we assign weights equal to the inverse of the squared uncertainties, $1/\sigma^2$, and do not mask out any of the low-SNR regions. We assume that the errors on the measured stellar flux values are dominated by photon noise, and we propagate them throughout the analysis \citep{wyttenbach2015}. The fractional errors are highest in the cores of the stellar lines (corresponding to low SNR). The pixels which contain overlapping planetary absorption and low-SNR stellar residuals therefore contribute very little to the overall transmission spectrum when combined in the planetary rest frame -- further information can be found in section~\ref{sec:method_velocity}. For comparison, we also calculate the combined transmission spectrum without including weights.

This work has three main goals. First, we aim to compare the weighted and unweighted approaches to combining the spectra and assess the impact on the resultant transmission spectrum. Next, we conduct a homogeneous survey of sodium absorption in ten gas giant atmospheres, applying a consistent method across all datasets to eliminate external influences from variations in analyses. Finally, we aim to look for trends linking sodium absorption to properties of the planet and system, with the ultimate goal of understanding how the characteristics of the atmospheres change over a diverse sample of gas giants. In section~\ref{sec:observations} we give an overview of the HARPS and HARPS-N observations used in this work. 
The data analysis steps for robustly extracting the transmission spectra are described in section~\ref{sec:transmission_spectra}.
In section~\ref{sec:results}, we confirm and report on Na detections in ten gas giant atmospheres, compare the approaches to combining the spectra, and discuss our results with reference to previous studies. We then use these results to search for trends, and discuss atmospheric heights of the sodium layer, sodium doublet line ratios, and atmospheric wind velocities in section~\ref{sec:trends}. Finally, a summary of the results and potential avenues for future research are presented in section~\ref{sec:conclusions}.

\section{Observations}
\label{sec:observations}

In this work, we use archival observations of the ten transiting exoplanet systems listed in Table~\ref{tab:planet_sample}, each of which has been observed either with the HARPS or HARPS-N spectrographs. HARPS is a fibre-fed, cross-dispersed echelle spectrograph installed on the ESO 3.6~m telescope in La Silla, Chile. 72 spectral orders are recorded over a range of 380--690~nm with a resolution of $R=115,000$. One spectral order is lost due to a gap in the two 4k~$\times$~4k pixel CCDs \citep{mayor2003}. The HARPS-N spectrograph is a similar instrument, with slightly different specifications and performance improvements (e.g. increased beam stability, increased reference precision, and improved image quality and quantum efficiency) made possible through updated components. The echelle spectrum is split into 69 orders and recorded on a single 4k~$\times$~4k pixel CCD, thus there is not a gap in the data as with HARPS. It is installed on the Telescopio Nazionale Galileo (TNG) in La Palma, Canary Islands \citep{cosentino2012}.

The instruments are housed in a vacuum enclosure within a temperature-controlled environment to reach the required stability and remove radial velocity drifts from variations in temperature, ambient air pressure, and humidity. They are fed by two fibres, which allow for simultaneous observing of the target on fibre A, and either the sky or calibration source on fibre B. 

The observing log in Table~\ref{tab:observing_log} summarises all of the observations used within this work -- all data are accessed through the ESO or TNG archives. Even when conducting a homogeneous survey, it is important to review the observations and remove any spectra that would introduce systematic errors or adversely affect the extracted transmission spectrum.
Further information can be found in sections \ref{sec:obs_wasp69b} to \ref{sec:obs_kelt9b}, where we highlight any data that were discarded from our analysis. 

All observations were automatically reduced with the HARPS Data Reduction Software (DRS) at the end of each exposure -- the DRS version used for each dataset is listed in Table~\ref{tab:observing_log}. Each spectral order is background subtracted, cosmic-ray corrected, flat-fielded, blaze corrected, and wavelength calibrated using the calibration frames taken at the beginning of each night. All orders are then merged and re-binned, giving a uniformly spaced (0.01~{\AA} wavelength resolution), one-dimensional spectrum in the Solar System barycentric rest frame -- we perform our analysis on this "s1d" product of the pipeline. The Fibre B column of Table~\ref{tab:observing_log} shows if the sky or calibration source was simultaneously observed alongside the target; for "Dark" observations, the pipeline uses the order location of fibre~B to perform CCD background correction.

It is ideal to obtain a number of exposures before and after the transit to create a master out-of-transit spectrum with high signal-to-noise. Most of the observations used within this work cover the full transit and a period shortly before and shortly after -- any exceptions to this are noted in the sections below. The spectra are defined as "fully out-of-transit", "fully in-transit", or "during ingress/egress" by modelling the orbit using the parameters listed in Tables~\ref{tab:parameters1-5} and \ref{tab:parameters6-10} in Appendix~\ref{app:system_parameters}.

Some planets in this sample have previously been observed with other spectrographs such as ESPRESSO \citep{pepe2013} and CARMENES \citep{quirrenbach2014}.
In this homogeneous study, we focus solely on data from HARPS and HARPS-N which share similar instrumental properties, helping to reduce the influence of systematics from different instruments and variations in the data reduction pipelines. Further investigation to include observations from other spectrographs may improve the quality of the transmission spectra, but is beyond the scope of this current work.

\begingroup
\renewcommand{\arraystretch}{1.2} 
\begin{table*}
    \centering
    \caption{Observing log for the HARPS and HARPS-N observations used within this work. All dates correspond to the starting date of the observing night (observations may occasionally start after 00:00 on the following day).
    $^\dagger$~Nights are labelled to refer to in the text without quoting the full date -- we use the first letter of the planet name and the first 1-3 numbers of the planet number (e.g. W121 for WASP-121b), and the corresponding numbered observing night (e.g. N1 for the first set of observations). 
    $^\ddagger$~The total number of observed spectra used within the analysis, and in brackets, the number of fully in- and fully out-of-transit spectra: total(in/out). We have not included discarded frames in these totals, or discarded nights of archival data in the full log -- further explanation regarding data rejection can be found in sections \ref{sec:obs_wasp69b} to \ref{sec:obs_kelt9b}. 
    *~All stellar spectra from this night were subtracted by their simultaneously observed sky spectra to remove telluric sodium. 
    $^\S$~Nights are numbered 1, 2, 3 to remain consistent with \citet{seidel2019}, despite W76-N4 having an earlier date than N2 and N3. 
    $^\P$~Nights are numbered 1 and 3 to remain consistent with \citet{cabot2020} and \citet{hoeijmakers2020}, accounting for the rejected second night.
    }
    \label{tab:observing_log}
    \begin{tabular*}{\textwidth}{l@{\extracolsep{\fill}}lllllll}
    \toprule
    Planet          & Date          & Night Name~$^\dagger$ & Instrument    & Program ID        & \# Spectra~$^\ddagger$& Fibre B       & DRS Version   \\
    \midrule
    WASP-69b        & 2016-06-04    & W69-N1                & HARPS-N       & CAT16A\_130       & 16(8/7)               & Sky~*         & HARPN\_3.7    \\
                    & 2016-08-04    & W69-N2                & HARPS-N       & CAT16A\_130       & 18(7/9)               & Sky~*         & HARPN\_3.7    \\
    HD~189733~b     & 2006-09-07    & H189-N1               & HARPS         & 072.C-0488(E)     & 20(10/8)              & Dark          & HARPS\_3.5    \\
                    & 2007-07-19    & H189-N2               & HARPS         & 079.C-0828(A)     & 39(18/20)             & Sky           & HARPS\_3.5    \\
                    & 2007-08-28    & H189-N3               & HARPS         & 079.C-0127(A)     & 40(18/20)             & Sky           & HARPS\_3.5    \\
    WASP-21b        & 2011-09-05    & W21-N1                & HARPS         & 087.C-0649(A)     & 22(12/8)              & Dark          & HARPS\_3.5    \\
                    & 2011-09-18    & W21-N2                & HARPS         & 087.C-0649(A)     & 19(12/5)              & Dark          & HARPS\_3.5    \\
                    & 2018-09-07    & W21-N3                & HARPS-N       & CAT18A\_D1        & 33(13/18)             & Sky~*         & HARPN\_3.7    \\
    WASP-49b        & 2015-12-06    & W49-N1                & HARPS         & 096.C-0331(B)     & 41(11/28)             & Sky           & HARPS\_3.8    \\
                    & 2015-12-31    & W49-N2                & HARPS         & 096.C-0331(B)     & 38(8/29)              & Sky           & HARPS\_3.8    \\
                    & 2016-01-14    & W49-N3                & HARPS         & 096.C-0331(B)     & 39(11/26)             & Sky           & HARPS\_3.8    \\
    WASP-79b        & 2012-11-12    & W79-N1                & HARPS         & 090.C-0540(H)     & 29(21/6)              & Dark          & HARPS\_3.5    \\
    WASP-76b        & 2012-11-11    & W76-N1~$^\S$          & HARPS         & 090.C-0540(F)     & 63(39/22)             & Dark          & HARPS\_3.5    \\
                    & 2017-10-24    & W76-N2~$^\S$          & HARPS         & 0100.C-0750(A)    & 49(26/21)             & Sky           & HARPS\_3.8    \\
                    & 2017-11-22    & W76-N3~$^\S$          & HARPS         & 0100.C-0750(A)    & 51(40/9)              & Sky           & HARPS\_3.8    \\
                    & 2017-08-16    & W76-N4                & HARPS         & 099.C-0898(A)     & 30(17/12)             & Sky~*         & HARPS\_3.8    \\
                    & 2018-09-02    & W76-N5                & HARPS         & 0101.C-0889(A)    & 39(21/17)             & Sky~*         & HARPS\_3.8    \\
    MASCARA-2b      & 2017-08-16    & M2-N1                 & HARPS-N       & CAT17A\_38        & 90(56/32)             & Sky           & HARPN\_3.7    \\
                    & 2018-07-12    & M2-N2                 & HARPS-N       & CAT18A\_34        & 108(54/52)            & Sky           & HARPN\_3.7    \\
                    & 2018-07-19    & M2-N3                 & HARPS-N       & CAT18A\_34        & 78(39/37)             & Sky           & HARPN\_3.7    \\
    WASP-121b       & 2017-12-31    & W121-N1~$^\P$         & HARPS         & 0100.C-0750(C)    & 35(14/19)             & Sky           & HARPS\_3.8    \\
                    & 2018-01-14    & W121-N3~$^\P$         & HARPS         & 0100.C-0750(C)    & 50(18/30)             & Sky           & HARPS\_3.8    \\
    WASP-189b       & 2019-04-14    & W189-N1               & HARPS         & 0103.C-0472(A)    & 126(67/58)            & Sky~*         & HARPS\_3.8    \\
                    & 2019-04-25    & W189-N2               & HARPS         & 0103.C-0472(A)    & 109(53/55)            & Sky           & HARPS\_3.8    \\
                    & 2019-05-06    & W189-N3               & HARPS-N       & CAT19A\_97        & 112(67/43)            & Sky           & HARPN\_3.7    \\
                    & 2019-05-14    & W189-N4               & HARPS         & 0103.C-0472(A)    & 122(65/55)            & Sky~*         & HARPS\_3.8    \\
    KELT-9b         & 2017-07-31    & K9-N1                 & HARPS-N       & A35DDT4           & 49(21/26)             & Sky           & HARPN\_3.7    \\
                    & 2018-07-20    & K9-N2                 & HARPS-N       & OPT18A\_38        & 46(22/23)             & Sky           & HARPN\_3.7    \\
    \bottomrule
    \end{tabular*}
\end{table*}
\endgroup

\subsection{WASP-69b}
\label{sec:obs_wasp69b}
Two full transits of WASP-69b \citep{anderson2014} were observed with the \mbox{HARPS-N} spectrograph from program CAT16A\_130. 
When compared to the other targets, the total number of spectra is low (16 for night W69-N1 and 18 for W69-N2) 
which may hinder our ability to identify planetary signals.
Using the same data, the transmission spectrum of \mbox{WASP-69b} was previously analysed by \citet{casasayas-barris2017} (to measure sodium absorption) and \citet{khalafinejad2021} (in combination with other low and high-resolution spectrographs).

\subsection{HD~189733~b}
\label{sec:obs_hd189733b}
Three transits of HD~189733~b \citep{bouchy2005} were observed with HARPS from programs 072.C-0488(E), 079.C-0828(A), and 079.C-0127(A). These observations have previously been used to detect and analyse sodium absorption in the upper atmosphere \citep[e.g.][]{wyttenbach2015, casasayas-barris2017, borsa2018, langeveld2021}. Observations on night H189-N2 started shortly after the ingress of the planet, so we are unable to include a before-transit sample of stellar spectra in our analysis. This may have implications on the extracted planetary signal for that night. Additional spectra were obtained on \mbox{2006-07-29}, however the second half of the transit was not observed due to bad weather conditions -- we therefore discarded these observations.

\subsection{WASP-21b}
\label{sec:obs_wasp21b}
Two transits of WASP-21b \citep{bouchy2010} were observed with HARPS (program 087.C-0649(A)) and one transit with HARPS-N (program CAT18A\_D1) -- these data were previously analysed to detect Na in the planetary atmosphere \citep{chen2020b}. The host star is the faintest target in this survey and multiple nights of data are needed to detect the sodium doublet with a good enough SNR.

Three more transits were observed with HARPS-N on 2018-10-03, 2018-10-16, and 2018-10-29 (program OPT18B\_42), however many exposures had very low SNR ($< 15$) at the centre of the 56th order (which contains the sodium doublet). After discarding the low-SNR frames, we were unable to extract a transmission spectrum that was not dominated by noise, and therefore chose not to include these nights in our final analysis.

\subsection{WASP-49b}
\label{sec:obs_wasp49b}
Three transits of WASP-49b \citep{lendl2012} were observed with HARPS from program 096.C-0331(B), and were previously analysed by \citet{wyttenbach2017} to detect sodium. From night \mbox{W49-N2}, we discarded seven spectra which had low SNR (<~20) at the centre of the 56th order: four out-of-transit and three in-transit.
These spectra had almost zero flux recorded in the stellar sodium line cores which adversely affected the transmission spectrum and introduced systematic errors (see section~\ref{sec:method_velocity}).
We also discarded one spectrum from \mbox{W49-N3} which contained noisy emission-like features, possibly due to twilight pollution.

\subsection{WASP-79b}
\label{sec:obs_wasp79b}
One transit of WASP-79b \citep{smalley2012} was observed with HARPS from program 090.C-0540(H). These observations have not previously been analysed for atmospheric detections with transmission spectroscopy methods, however they have been used to measure the RM effect and analyse the orbital geometry \citep{brown2017}.

The second exposure obtained at 03:42 (which captured the transit ingress) was discarded due to a contaminating emission-like feature in the sodium doublet, possibly due to telluric sodium. The sky spectra are not available to check or correct this. The stellar apparent magnitude of 10.044 makes WASP-79b one of the fainter targets in our sample, and with only one observed transit, our ability to extract a transmission spectrum with good SNR is limited. However, sodium is one of the strongest features within this wavelength domain.

\subsection{WASP-76b}
\label{sec:obs_wasp76b}
The three transits of WASP-76b \citep{west2016} from HARPS programs 090.C-0540(F) and 0100.C-0750(A) have previously been analysed to show the presence of sodium with broadened line profiles \citep{seidel2019, zak2019}, study atmospheric winds using the sodium doublet \citep{seidel2021}, and detect an asymmetric signature of iron absorption \citep[e.g. see][for both ESPRESSO and HARPS data]{ehrenreich2020, kesseli2021}. 
The final 15 exposures obtained between 04:58 and 06:18 on night W76-N3 were discarded due to cloud cover \citep[for further discussion, see][]{seidel2019}. 
The first exposure starting at 23:44 on night \mbox{W76-N1} was also removed due to abnormal telluric emission-like features around the sodium doublet.

In this work, we incorporate two additional transits (nights \mbox{W76-N4} and W76-N5) obtained from programs 099.C-0898(A) and \mbox{0101.C-0889(A)}, which have not been included in previous studies. On night W76-N4, we discarded the final two exposures due to telluric sodium emissions which were unable to be corrected by subtracting the sky spectra. The observations on this night also stopped shortly before the transit egress, thus no after-transit spectra were obtained. One further transit was observed with HARPS-N on \mbox{2017-10-26} (GAPS programme), however there were only six fully out-of-transit spectra. We were unable to extract a good-quality transmission spectrum with an adequate SNR, and therefore chose to discard these observations.

\subsection{MASCARA-2b/KELT-20b}
\label{sec:obs_mascara2b}
MASCARA-2b \citep{talens2018}, also known as KELT-20b \citep{lund2017}, orbits a bright and rapidly rotating star \mbox{($v\sin{i} = 114\pm3$~km~s$^{-1}$)}. Three full transits were observed with \mbox{HARPS-N} under programs CAT17A\_38 and CAT18A\_34, and the data have previously been used to detect sodium \citep{casasayas-barris2018} and a number of other atomic species \citep{casasayas-barris2019, nugroho2020, stangret2020}.

From night M2-N2, we discarded eight pre-transit exposures that had a SNR lower than 53 at the centre of the 56th order. The SNR of other exposures on the same night was between 58 and 112. The sample of 52 remaining out-of-transit exposures is large enough to build a good-quality master-out spectrum (see section~\ref{sec:method_velocity}).

\subsection{WASP-121b}
\label{sec:obs_wasp121b}
Two transits of WASP-121b \citep{delrez2016} were observed with HARPS from program 0100.C-0750(C), and we label these nights W121-N1 and W121-N3. These observations have been analysed in previous studies to detect a number of chemical species including sodium \citep{ben-yami2020, cabot2020, hoeijmakers2020}, and analyse the RM effect and atmospheric structure \citep{bourrier2020}. We ignored an additional night of data obtained on \mbox{2018-01-09} (which would be referred to as \mbox{W121-N2}) due to low-SNR of the in-transit spectra. To verify the exclusion, we included this night in a separate analysis and found that it did not change the results significantly when using the weighted average approach.

\subsection{WASP-189b}
\label{sec:obs_wasp189b}
WASP-189b \citep{anderson2018} occupies a polar orbit around a rapidly rotating ($v\sin{i} = 97.1\pm2.1$~km~s$^{-1}$) host star, and is the brightest target in this survey. The high SNR makes this planet an excellent target for atmospheric characterisation. Three transits were observed with HARPS from program 0103.C-0472(A), and one transit with HARPS-N (program CAT19A\_97). These data have recently been analysed to detect multiple chemical species in the planetary atmosphere \citep{stangret2021, prinoth2022}.

Observations on night W189-N2 started shortly after the transit ingress and there are no before-transit spectra. This is not the ideal situation (where observations would cover the full transit and a period shortly before and after), however most of the transit is observed and there are enough in- and out-of-transit exposures to use for extracting the transmission spectrum. Archival data obtained on 2018-03-26 from program 0100.C-0847(A) can also be accessed from the HARPS archive. However, we chose not to include this night in our analysis because a significant portion of the transit was not observed.

\subsection{KELT-9b}
\label{sec:obs_kelt9b}
KELT-9b is an ultra-hot Jupiter with one of the hottest known planetary equilibrium temperatures, undergoing extreme UV irradiation by its rapidly rotating ($v\sin{i} = 111.40\pm1.27$~km~s$^{-1}$) host star \citep{gaudi2017}. Two full transits were observed with HARPS-N (programs A35DDT4 and OPT18A\_38). These observations were previously used to detect a number of atomic species in the planetary atmosphere with the cross-correlation method \citep[e.g.][]{hoeijmakers2018, hoeijmakers2019, yan2019, wyttenbach2020}.

\section{Transmission spectra}
\label{sec:transmission_spectra}

To calculate the planetary transmission spectrum from HARPS and HARPS-N observations, we primarily follow the methodology outlined in \citet{langeveld2021} and references therein. This is briefly summarised in the following sections, along with any additional considerations.

\begin{figure*}
    \centering
    \includegraphics[width=\textwidth]{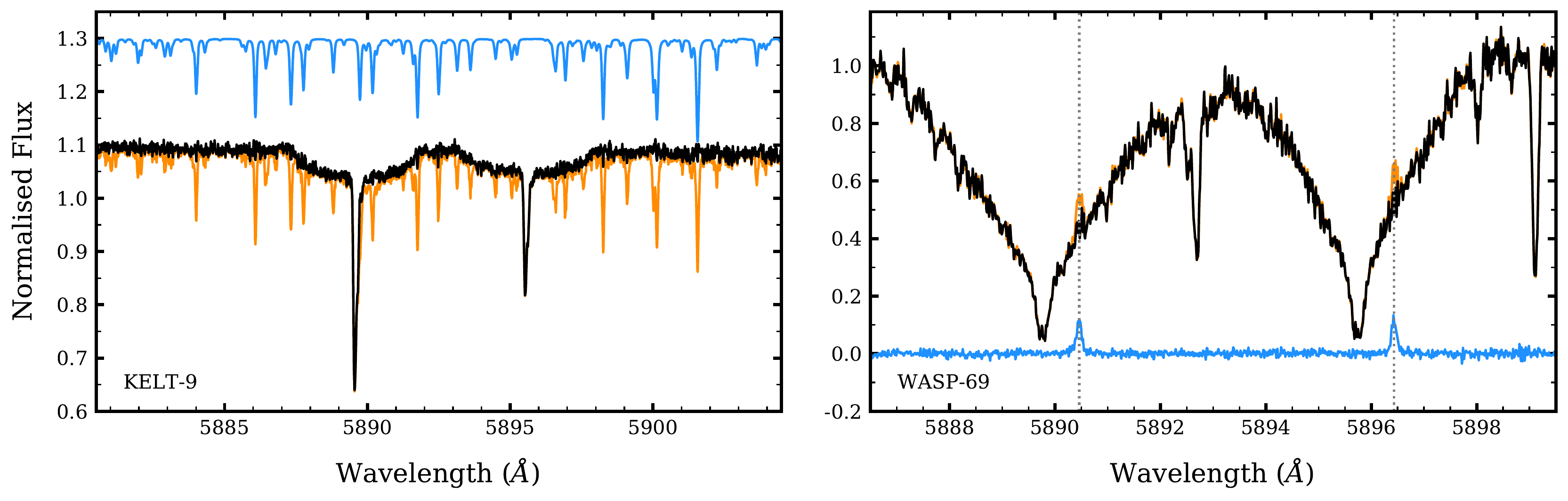}
    \caption{Examples of telluric contamination within the sodium region of the observed spectra. \textit{Left panel:} a normalised KELT-9 spectrum which contains strong telluric lines (orange), together with the model of telluric water and oxygen absorption from \texttt{molecfit} (blue) -- spectra are plotted with a y-axis offset for clarity. After dividing the stellar spectrum by the model, all telluric lines are reduced down to the noise level (black). \textit{Right panel:} a spectrum of WASP-69 observed with fibre A of the HARPS spectrograph (orange), and the simultaneously observed sky spectrum from fibre B (blue). The location of the sodium doublet in Earth's rest frame is indicated by the dotted line, and noticeable sodium emission can be seen in these regions. Subtracting the sky spectrum (blue) from the observed spectrum (orange) removes the emission features (black).}
    \label{fig:telluric_sky}
\end{figure*}

\begin{figure*}
    \centering
    \includegraphics[width=\textwidth]{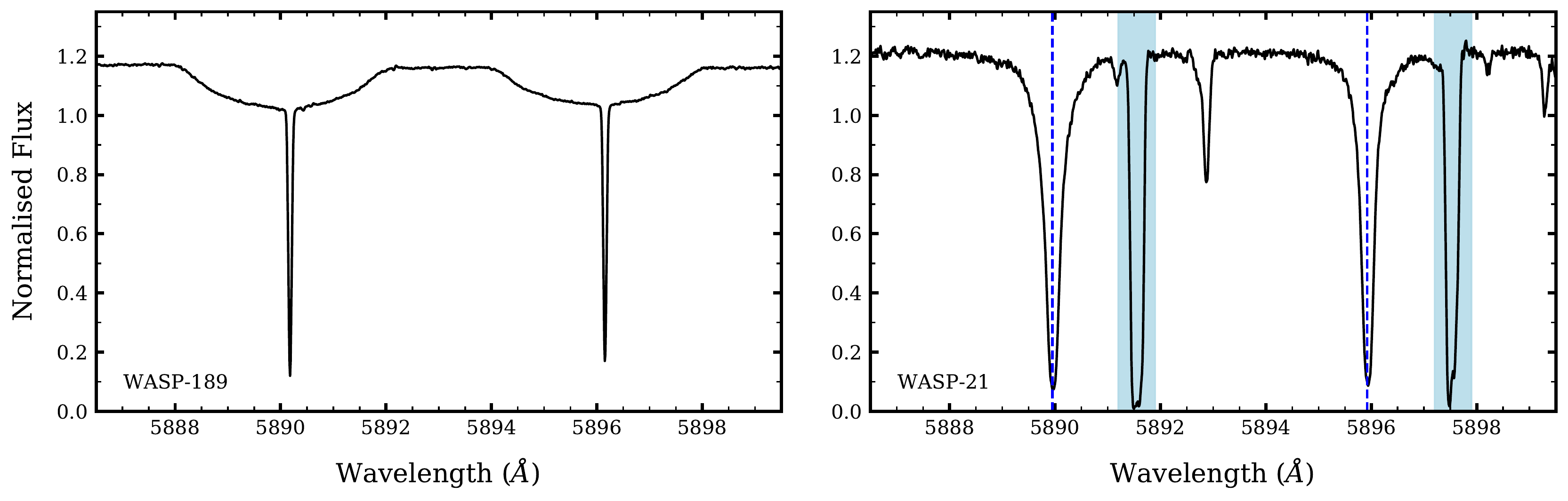}
    \caption{Examples of extra sodium absorption, possibly due to ISM contaminants. \textit{Left panel:} the master out-of-transit spectrum for WASP-189b (night W189-N3). Deep and narrow features that are slightly offset from the centre of the stellar lines are characteristic of ISM absorption. \textit{Right panel:} the master out-of-transit spectrum for WASP-21b (night W21-N3). The light-blue shaded regions highlight strong ISM absorption features which reduce the flux to near zero -- these regions were masked out in the analysis, which is possible due to the significant offset from the location of the sodium doublet in the stellar rest frame (blue dashed lines).}
    \label{fig:ISM_contamination}
\end{figure*}

\subsection{Telluric and interstellar contamination}
\label{sec:method_tellurics}

First, we limit the spectral range to 4000--6800~{\AA} (from 3781--6912~{\AA}) to reduce systematics from low throughput or strong telluric contamination at the edges of the full spectrum, and apply the same cleaning and normalisation processes as in previous work. 
A model of Earth's atmospheric absorption for each observed spectrum was produced using \texttt{molecfit} v1.2.0 \citep{smette2015, kausch2015}, which was previously shown to be robust at reducing telluric effects in this wavelength range over nights with varying observing conditions \citep{langeveld2021}. 
\texttt{Molecfit} is an ESO tool specifically designed for this purpose, which uses a line-by-line radiative transfer model of the Earth's atmosphere to fit synthetic transmission spectra to the observed data.

The observed stellar spectra are first shifted from the Solar System barycentric rest frame to the telescope rest frame using the Barycentric Earth Radial Velocity (BERV) values stored within the HARPS s1d file headers. We follow the steps outlined in previous work to select 15--20 small ($<2$~{\AA}) regions of isolated telluric H$_2$O and O$_2$ lines for each night, and provide \texttt{molecfit} with the same input settings that were used for correcting observations of HD~189733~b \citep[see][]{allart2017, langeveld2021}. The output from \texttt{molecfit} gives a unique set of fit parameters for each observed spectrum, which are read by the \texttt{calctrans} tool to fit the atmospheric model to the full-resolution data. Each spectrum is then divided by its respective telluric model to remove the contamination down the noise level.
The left-hand panel of Figure~\ref{fig:telluric_sky} shows a comparison of a telluric corrected (black) and uncorrected (orange) KELT-9b spectrum for night \mbox{K9-N1}. By comparison with the \texttt{molecfit} model (blue), it can clearly be seen that the telluric contamination is reduced to the continuum level without significantly changing other parts of the stellar spectrum.

Evidence of telluric sodium was also present in some nights of data. Where possible, we compared all stellar spectra to their simultaneously observed sky spectra to check for telluric sodium emission features. We found noticeable emission features in nights W21-N3, W69-N1, W69-N2, W76-N4, W76-N5, W189-N1, and W189-N4 -- an example of a spectrum from W69-N1 is shown in the right-hand panel of Figure~\ref{fig:telluric_sky}. For these nights, we subtracted the observed sky spectra (blue) from the stellar spectra (orange) to remove the telluric sodium features before proceeding with the analysis. Direct subtraction of the sky spectra may not account for differences in the efficiency of the two fibres. However, when inspecting the spectra, we found the corrections to be adequate to the noise level of the continuum and the fibre efficiencies were not considered.

The observations of some targets from the sample exhibited deep, narrow features characteristic of interstellar sodium absorption. Absorption lines are imprinted within the stellar spectra at a slight offset to the rest frame location of the sodium doublet -- the wavelength offset depends on the radial velocity of the interstellar absorber. In most cases \citep[e.g. MASCARA-2b:][]{casasayas-barris2018}, the interstellar absorption does not vary significantly over the observing period and is removed when dividing each stellar spectrum by the master-out spectrum. The left-hand panel of Figure~\ref{fig:ISM_contamination} shows the master-out spectrum for WASP-189b (night W189-N3). The deep and narrow absorption lines (at $\sim5890.2$~{\AA} and $\sim5896.2$~{\AA}) that are imprinted within the broad stellar sodium lines are evidence of the interstellar absorption. Similar artefacts can also be seen for KELT-9b in the left-hand panel of Figure~\ref{fig:telluric_sky}.

Observations for WASP-21b posed a larger problem: the interstellar sodium absorption was much stronger and reduced the flux to near zero. As a result, the SNR in these regions is much lower than the rest of the spectrum which drastically affects the quality of the combined planetary transmission spectrum. The right-hand panel of Figure~\ref{fig:ISM_contamination} shows the master-out spectrum for WASP-21b (night \mbox{W21-N3}). Strong interstellar sodium absorption can be seen within the blue shaded regions, causing the measured flux to drop to near zero. Fortunately, the radial velocity of the contaminating source was large enough to offset the absorption completely from the stellar lines (and thus the location of the planetary sodium absorption). We were therefore able to mask out the shaded regions without affecting the final planetary transmission spectrum, similarly to \citet{chen2020b}.

\subsection{Velocity corrections}
\label{sec:method_velocity}
Doppler shifts due to stellar reflex motion, systemic velocity, and planetary radial velocity can impact the quality of the planetary transmission spectrum and must be accounted for. RM and CLV effects present an additional problem, and are discussed in section~\ref{sec:method_CLVRM}.

First, we model the stellar reflex motion and systemic velocity assuming a circular orbit: 
\begin{equation}
    v_{\ast} = K_{\ast}\sin{(2\pi\phi)} +v_{\text{sys}} ~,
    \label{eqn:stellar_RV}
\end{equation}
where $K_{\ast}$ is the stellar radial velocity semi-amplitude, $\phi$ is the phase and $v_{\text{sys}}$ is the systemic velocity. Each stellar spectrum is Doppler shifted using the modelled $v_{\ast}$ value and linearly interpolated back to the uniform 0.01~{\AA} grid. We chose not to perform the shift using the HARPS measured stellar radial velocities which should include the "RM anomaly" offset during the transit \citep{rossiter1924, mclaughlin1924, gaudi2007, digloria2015, casasayas-barris2017, triaud2018}; for the fainter or faster-rotating stars in this survey, there are inconsistencies in the radial velocity measurements (e.g. velocities that are much higher or much lower than the others) which would induce more error than the actual effect.

For targets with additional interstellar medium (ISM) sodium contamination, Doppler shifting to correct for stellar radial velocity may prevent the ISM absorption from being fully removed when the in-transit spectra are divided by the master-out spectrum. This was the case for KELT-9b and WASP-189b, and minimally for MASCARA-2b (WASP-21b is treated separately). The maximum Doppler shift during the transit due to the stellar radial velocity varies from star to star, but is consistently less than one wavelength increment (0.01~{\AA}) -- often around 0.001--0.003~{\AA} (50--150~m~s$^{-1}$) in the sodium region. This is small in comparison to the systemic velocity and planetary radial velocity, and should not significantly impact our results. For fast-rotating stars with broadened stellar lines, the radial velocity shift does not have a significant effect on the planetary transmission spectrum and can be ignored \citep{casasayas-barris2018, casasayas-barris2019, cabot2021}. However, since we are conducting a homogeneous survey using a variety of data, we opt to correct for the stellar radial velocity to keep the method consistent for all targets. If ISM contamination is not fully removed, the residual spectra (equation \ref{eqn:transmission_residual}) may contain a trail of low-SNR features which are falsely enhanced/decreased, similar to those created by deep stellar lines. When the spectra are shifted into the planetary rest frame and combined, the low-SNR residuals are no longer stacked at the same wavelength. To check the magnitude of this effect for KELT-9b, WASP-189b, and MASCARA-2b, we performed the same analysis without correcting for the stellar reflex motion and the results were consistent within 1$\sigma$.

Similarly, the planetary radial velocity was calculated for all observed phases using 
\begin{equation}
    v_{\text{p}} = K_{\text{p}}\sin{(2\pi\phi)} ~,
    \label{eqn:planet_RV}
\end{equation}
where $K_{\text{p}}$ is the planetary radial velocity semi-amplitude, derived through the relationship $K_{\text{p}} = -K_{\ast}(M_{\ast}/M_{\text{p}})$.

For each night, we combine all fully out-of-transit spectra to form a master-out spectrum, $\overline{f}_{\text{out}}(\lambda)$. The corrected stellar spectra, $f(\lambda,t)$, are divided by the master-out, and hereafter referred to as residual spectra: 
\begin{equation}
    \Re(\lambda,t) = \frac{f(\lambda,t)}{\overline{f}_{\text{out}}(\lambda)} ~,
    \label{eqn:transmission_residual}
\end{equation}
It is useful to compare the residual spectra for both in-transit and out-of-transit observations to check for false signals, CLV and RM effects, and low-SNR residuals from deep stellar line cores.

Effects from instrumental systematics and weather variations are removed by continuum normalising each residual spectrum using a third-order polynomial. We then Doppler shift them into the planetary rest frame according to the calculated $v_\text{p}$, and linearly interpolate back to the 0.01~{\AA} grid. At this stage, we divide each residual spectrum by its corresponding CLV and RM model (produced via the steps outlined in section~\ref{sec:method_CLVRM}) to remove false signals which arise from the transit of a planet across a rotating star and not from its atmosphere. Finally, the combined transmission spectrum is computed using fully in-transit residual spectra:
\begin{equation}
    {\Re'}(\lambda) = \sum_{t_{\text{in}}}\Re(\lambda,t_{\text{in}})|_{v_{\text{p}}(t_{\text{in}})} - 1 = \frac{F_{\lambda, \text{in}}}{F_{\lambda, \text{out}}} -1 ~,
    \label{eqn:transmission_combined}
\end{equation}
where $F_{\lambda, \text{in}}$ is the Doppler-corrected combined in-transit spectrum \citep{brown2001, wyttenbach2015, langeveld2021}. We ignore ingress/egress exposures for all targets in this survey to prevent inclusion of partially out-of-transit data in the in-transit sample.

The continuum of the resultant transmission spectrum sometimes contains noticeable broadband variations which do not originate from the astrophysical source. This could possibly be due to imperfect blaze correction, imperfect colour correction, or problems with the atmospheric dispersion corrector \citep[e.g. for MASCARA-2b, see][]{casasayas-barris2019}. A median filter with a width of 15.01~{\AA} was applied to remove these broad variations without affecting the narrow planetary signals.

\begin{figure}
    \centering
    \includegraphics[width=\columnwidth]{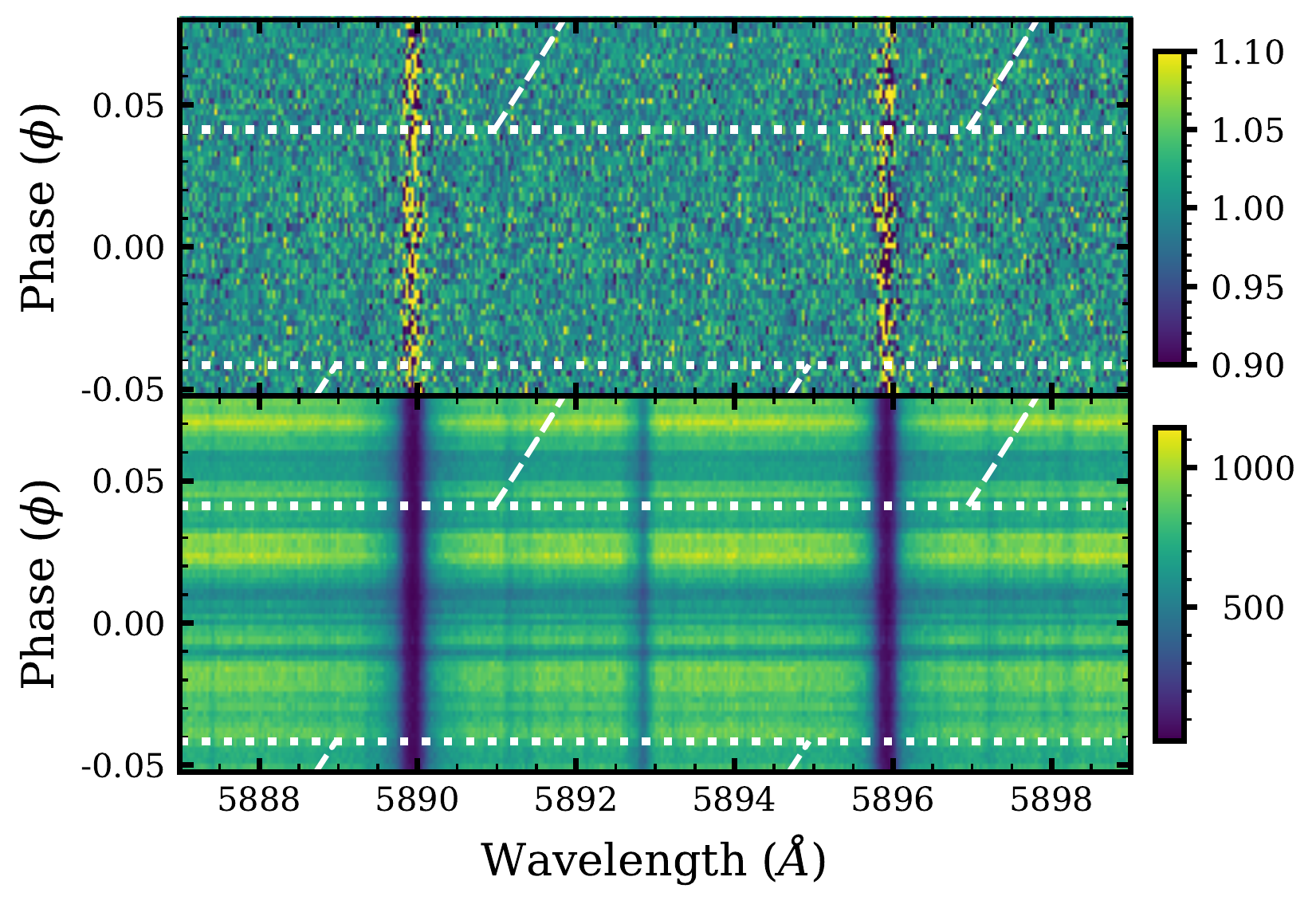}
    \caption{\textit{Top panel:} residual spectra for WASP-76b (night \mbox{W76-N1}), after dividing all stellar spectra by the master-out spectrum. \textit{Bottom panel:} $1/\sigma^2$ weights associated with the residual spectra. Vertical bands of low-SNR residuals can be seen in the top panel around 5890 and 5896~{\AA}, and the corresponding weights are very low. "Fully-in-transit" spectra are located between the horizontal white dotted lines, and the path traced out by sodium in the planetary rest frame is indicated by the white dashed lines -- spectra are combined along this path, so the low-SNR residuals are no longer stacked at the same wavelengths.}
    \label{fig:residuals_weights}
\end{figure}

We define "combining the spectra" as finding the average -- either using the simple or weighted mean. We aim to analyse the differences between combining the spectra with and without weights, and therefore repeat the above calculations twice: once using the simple average, and once using the inverse of the squared uncertainties $(1/\sigma^2)$ as weights. This process affects the master-out spectrum and the combined transmission spectrum. The errors of the measured stellar flux values are assumed to be dominated by photon noise, and the Poisson uncertainties are propagated throughout this analysis. Figure~\ref{fig:residuals_weights} shows an example of the residual spectra for WASP-76b (W76-N1) and their corresponding weights -- all spectra between the horizontal white dotted lines are defined as "fully in-transit", and the white dashed lines trace out the location of the sodium lines in the planetary rest frame. As can be seen, there are two vertical bands of low-SNR residuals in the stellar rest frame -- if the spectra are combined in the planetary rest frame using the weighted average, the noisy residuals will contribute significantly less to the overall transmission spectrum.

On a similar note, we chose to divide the CLV and RM models before combining the in-transit residual spectra, which removes the false signals before considering weighted averaging. An alternative method may involve dividing the final transmission spectrum by a combined CLV and RM model -- this would produce the same result provided that the weights used to combine the data are also used to combine the models (if they are chosen to be included in the average).

\subsection{Correcting CLV and RM effects}
\label{sec:method_CLVRM}
The combined CLV and RM effects are negligible for some planets in this survey (e.g. those with slowly rotating host stars). However, there are certain scenarios where it is vital to make this correction otherwise the resulting artefacts may be of greater amplitude than the planetary absorption.

We model combined CLV and RM effects on the transmission spectra identically to \citet{cabot2020}, whose approach was based on \citet{yan2017} and \citet{casasayas-barris2019}. In summary, the star is represented on an 80~$\times~$80 pixel grid. Pixels within the star's radius are assigned a high-resolution synthetic spectrum based on Kurucz \texttt{ATLAS9} stellar atmosphere models. The synthetic spectra are generated with appropriate stellar parameters and across a range of $\mu$-values, which allows us to account for limb darkening via interpolation. We also Doppler shift each pixel's spectrum by the local radial velocity of the rotating stellar surface. Pixels outside the star's radius have zero flux. We simulate the transit by determining the position of an opaque disc representing the planet, setting the occulted stellar pixels to zero flux, and integrating the stellar spectrum. This procedure is repeated for all observed phases/timestamps. Transmission spectra are calculated according to equation~\ref{eqn:transmission_combined}, and are divided out of the observed spectra. Adopted values for orbital and stellar parameters can be found in Tables~\ref{tab:parameters1-5} and \ref{tab:parameters6-10}, along with the associated references. An example of the RM/CLV correction for WASP-189b is shown in Figure~\ref{fig:rmclv_correction} in Appendix~\ref{app:rmclv}. Additionally, we have included examples of models for each planet in Figure~\ref{fig:rmclv_models}, showing the diversity of artefacts from different orbital geometries.

We note that the model does not always perfectly capture the observed RM/CLV artefacts; in some cases (e.g. WASP-189b) the model appeared slightly offset, or had a lower amplitude. These discrepancies likely arise from uncertainties in the orbital parameters (for example, slight variations in inclination $i$ effectively shift the model under certain transit geometries), uncertainty in $R_{\text{p}}/R_*$, and wavelength dependence of $R_{\text{p}}/R_*$. In order to improve the efficacy of the correction, we performed a chi-squared minimisation fit of the RM/CLV model to the data, allowing for both a constant velocity offset and multiplicative scale factor. This fit approximates variations in the model that would arise from slight perturbations to the input parameters, and is sufficient for removing RM/CLV artefacts. However, it may be useful in future work to explore constraints with a Markov Chain Monte Carlo (MCMC) analysis involving all input parameters and the full forward model described above.

Additionally, the sky projected obliquity measurements for \mbox{WASP-21b} and WASP-49b (see Table~\ref{tab:parameters1-5}) both have large uncertainties which could drastically change the shape of the RM/CLV model, and thus the shape of the absorption features in the transmission spectrum. However, we found that the combined effect of the models was negligible compared to the noise level of the transmission spectrum, which was also confirmed by previous work using similar methods \citep{chen2020b, wyttenbach2017}. We also do not expect the RM effect to be significant for the slowly rotating host stars.

\begin{figure*}
    \centering
    \includegraphics[width=0.99\textwidth]{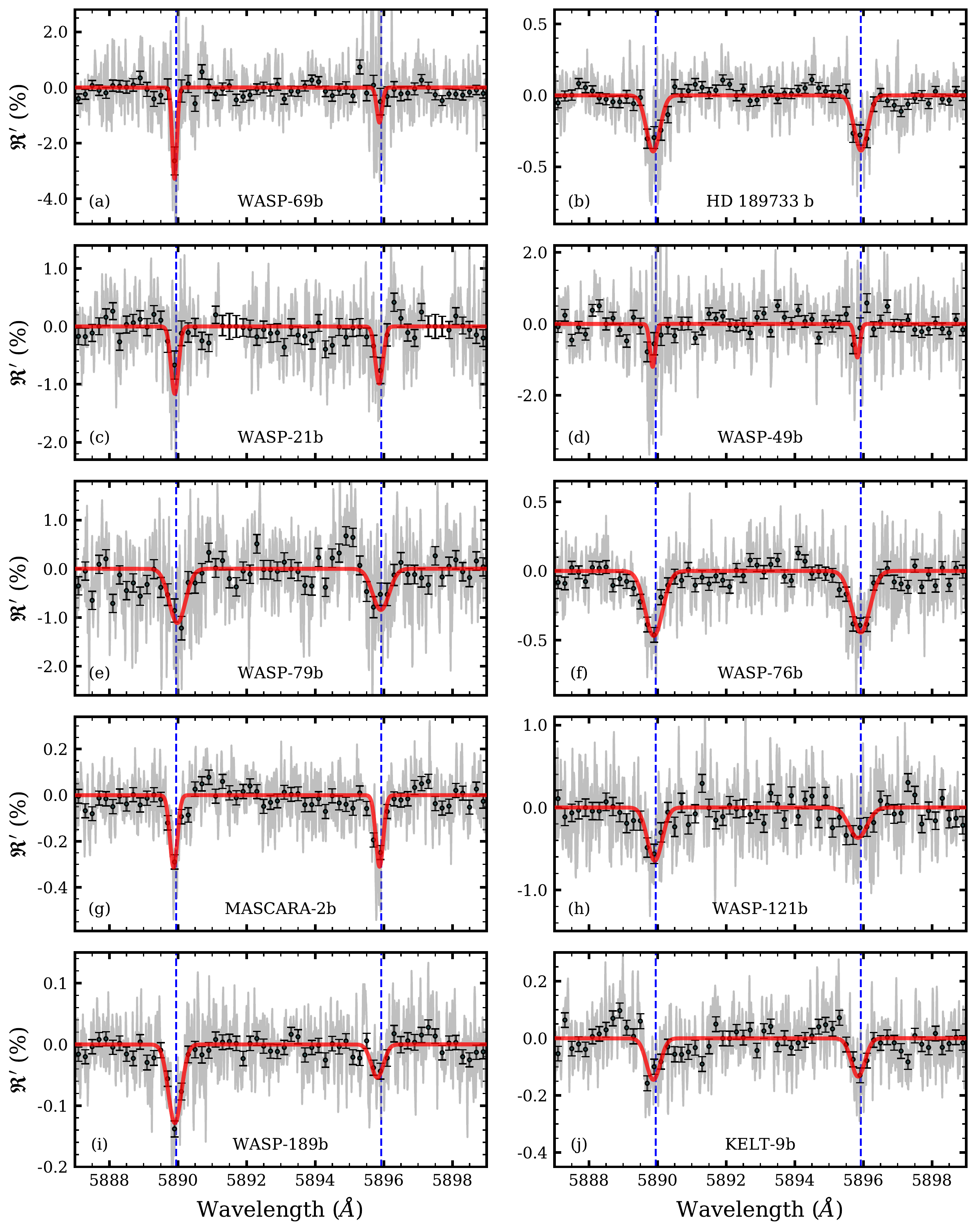}
    \caption{Evidence of sodium absorption in the atmospheres of ten transiting gas giant planets. Each panel shows the weight-combined transmission spectrum for the respective (labelled) planet: full-resolution data in grey, and binned 20$\times$ in black. Gaussian profiles (red) are fitted to the full-resolution data, and the rest frame positions of the sodium D-lines are indicated with blue dashed lines. Note: the y-axis scale changes for each panel to clearly show the absorption features.}
    \label{fig:ten_transmission}
\end{figure*}

\section{Results: A homogeneous sodium survey}
\label{sec:results}
In this section, we summarise detections of sodium in the ten gas giant exoplanets chosen for this survey, compare the weighted and unweighted approaches for combining the data, and discuss reproductions of previous work.

\subsection{Detections of sodium}
\label{sec:Na_detections}

The final combined transmission spectra were initially binned to a 0.2~{\AA} resolution (20 points per bin) for visual inspection. Gaussian profiles were fitted to the full-resolution data using the \texttt{LevMarLSQFitter} module from \texttt{astropy}, accounting for errors that were propagated from the photon noise of the observed spectra. Three parameters were fitted and measured for each line: the depth or line contrast $(\mathcal{D})$; centroid $(\lambda_0)$; and the full width at half maximum (FWHM). As defined by equation~\ref{eqn:transmission_combined}, absorption features extend below the continuum and are negative in value. Gaussian profiles provide a good approximation for the line cores and is a standard approach in previous work. Other models that account for atmospheric effects can be used \citep[e.g.][]{ehrenreich2006, wyttenbach2015, wyttenbach2020, pino2018, oza2019, gebek2020, hoeijmakers2020, seidel2020a, seidel2021}, but this is beyond the scope of the current work. 

Figure~\ref{fig:ten_transmission} shows the weight-combined transmission spectra for all ten gas giants analysed in this work, ordered according to their equilibrium temperature. Gaussian profiles (red) are fitted to the full-resolution data (grey) as described above, and the binned points are shown in black. We confirm the previously reported detections of Na in WASP-69b \citep{casasayas-barris2017}, HD~189733~b \citep{wyttenbach2015}, WASP-21b \citep{chen2020b}, WASP-49b \citep{wyttenbach2017}, WASP-76b \citep{seidel2019, zak2019}, MASCARA-2b \citep{casasayas-barris2019}, WASP-121b \citep{cabot2020, hoeijmakers2020}, WASP-189b \citep{prinoth2022}, and KELT-9b \citep{hoeijmakers2019}. These results are discussed further in sections~\ref{sec:results_comparison} and \ref{sec:results_previous_work}, along with the same measurements for the unweighted spectra.

We report on a new detection of Na in the atmosphere of \mbox{WASP-79b} -- shown in panel (e) of Figure~\ref{fig:ten_transmission}. From the Gaussian fits, the measured line contrasts were $-1.12 \pm 0.23~\%$ (D2) and $-0.85 \pm 0.22~\%$ (D1), averaging to $-0.98 \pm 0.16~\%$. 
The stellar spectral lines were broadened due to rotation, thus the residual transmission spectra did not contain narrow bands of low-SNR regions from deep line cores (see Figure~\ref{fig:residuals_weights}).
However, the host star is one of the fainter targets in our sample and there is only one night of data with few out-of-transit frames. Further observations of \mbox{WASP-79b} transits would be beneficial for constraining the Gaussian fit parameters and improving the SNR.

\begingroup
\renewcommand{\arraystretch}{1.2} 
\begin{table*}
    \centering
    \caption{Measured depths of the Gaussian fits to the sodium doublet lines for all planets in this survey, with comparison to previous studies. 
    The "--" symbol is used to denote no previous measurements of individual lines via Gaussian fits, or if the combination of data have not been analysed together before. 
    In addition to our homogeneous analysis, we also measured the line depths for WASP-76b and MASCARA-2b using the same observations/nights as the referenced work for direct comparison.
    $^\dagger$~Results for one MASCARA-2b transit (M2-N1). 
    $^\ddagger$~Results for three MASCARA-2b transits (M2-N1, M2-N2 and M2-N3). 
    $^\S$~Results for three WASP-76b transits (W76-N1, W76-N2 and W76-N3). 
    *~Results for five WASP-76b transits (W76-N1, W76-N2, W76-N3, W76-N4 and W76-N5). 
    References to previously published results: 
    (1)~\citet{casasayas-barris2017}, 
    (2)~\citet{wyttenbach2015}, 
    (3)~\citet{langeveld2021}, 
    (4)~\citet{chen2020b}, 
    (5)~\citet{wyttenbach2017}, 
    (6)~\citet{seidel2019}, 
    (7)~\citet{zak2019}, 
    (8)~\citet{casasayas-barris2018}, 
    (9)~\citet{casasayas-barris2019}, 
    (10)~\citet{cabot2020}, 
    (11)~\citet{hoeijmakers2020}, 
    (12)~\citet{prinoth2022}, 
    (13)~\citet{hoeijmakers2019}.
    }
    \label{tab:all_results}
    \begin{tabular*}{\textwidth}{l@{\extracolsep{\fill}}lclllllllll}
    \toprule
    \multirow{2}{*}{Planet} && \multirow{2}{*}{Ref.} && \multicolumn{2}{c}{Value from reference}        && \multicolumn{2}{c}{This work (weighted)}        && \multicolumn{2}{c}{This work (unweighted)}    \\
                            &&   && \multicolumn{1}{c}{D2~(\%)} & \multicolumn{1}{c}{D1~(\%)} && \multicolumn{1}{c}{D2~(\%)} & \multicolumn{1}{c}{D1~(\%)} && \multicolumn{1}{c}{D2~(\%)} & \multicolumn{1}{c}{D1~(\%)} \\
    \midrule
    WASP-69b                && 1                     && $-5.80 \pm 0.30$  & \multicolumn{1}{c}{--}      && $-3.28 \pm 0.70$  & $-1.26 \pm 0.61$            && $-3.64 \pm 0.82$  & $-1.19 \pm 0.80$  \\
    HD~189733~b             && 2                     && $-0.64 \pm 0.07$  & $-0.40 \pm 0.07$            && $-0.39 \pm 0.06$  & $-0.39 \pm 0.06$            && $-0.57 \pm 0.07$  & $-0.47 \pm 0.07$  \\
                            && 1                     && $-0.72 \pm 0.05$  & $-0.51 \pm 0.05$            &&                   &                             &&                   &                   \\
                            && 3                     && $-0.64 \pm 0.07$  & $-0.53 \pm 0.07$            &&                   &                             &&                   &                   \\
    WASP-21b                && 4                     && $-1.18 \pm 0.24$  & $-0.84 \pm 0.17$            && $-1.16 \pm 0.22$  & $-0.99 \pm 0.20$            && $-1.17 \pm 0.20$  & $-0.96 \pm 0.19$  \\
    WASP-49b                && 5                     && $-1.99 \pm 0.49$  & $-1.83 \pm 0.65$            && $-1.20 \pm 0.46$  & $-0.94 \pm 0.42$            && $-1.93 \pm 0.51$  & $-2.01 \pm 0.47$  \\
    WASP-79b                && --                    && \multicolumn{1}{c}{--} & \multicolumn{1}{c}{--} && $-1.12 \pm 0.23$  & $-0.85 \pm 0.22$            && $-1.14 \pm 0.23$  & $-0.72 \pm 0.23$  \\
    WASP-76b~$^\S$          && 6                     && $-0.33 \pm 0.09$  & $-0.51 \pm 0.08$            && $-0.52 \pm 0.06$  & $-0.55 \pm 0.06$            && $-0.44 \pm 0.10$  & $-0.61 \pm 0.08$  \\
                            && 7                     && $-0.57 \pm 0.08$  & $-0.65 \pm 0.07$            &&                   &                             &&                   &                   \\
    WASP-76b~*              && --                    && \multicolumn{1}{c}{--} & \multicolumn{1}{c}{--} && $-0.47 \pm 0.05$  & $-0.45 \pm 0.05$            && $-0.44 \pm 0.08$  & $-0.49 \pm 0.07$  \\
    MASCARA-2b~$^\dagger$   && 8                     && $-0.44 \pm 0.11$  & $-0.37 \pm 0.08$            && $-0.37 \pm 0.09$  & $-0.40 \pm 0.07$            && $-0.34 \pm 0.09$  & $-0.38 \pm 0.07$  \\
    MASCARA-2b~$^\ddagger$  && 9                     && $-0.33 \pm 0.06$  & $-0.35 \pm 0.05$            && $-0.31 \pm 0.04$  & $-0.31 \pm 0.04$            && $-0.31 \pm 0.04$  & $-0.31 \pm 0.04$  \\
    WASP-121b               && 10                    && $-0.69 \pm 0.12$  & $-0.25 \pm 0.09$            && $-0.65 \pm 0.10$  & $-0.37 \pm 0.09$            && $-0.54 \pm 0.11$  & $-0.28 \pm 0.09$  \\
                            && 11                    && $-0.56 \pm 0.07$  & $-0.27 \pm 0.07$            &&                   &                             &&                   &                   \\
    WASP-189b               && 12                    && \multicolumn{1}{c}{--} & \multicolumn{1}{c}{--} && $-0.13 \pm 0.01$  & $-0.05 \pm 0.01$            && $-0.14 \pm 0.01$  & $-0.07 \pm 0.01$  \\
    KELT-9b                 && 13                    && \multicolumn{1}{c}{--} & \multicolumn{1}{c}{--} && $-0.16 \pm 0.03$  & $-0.16 \pm 0.03$            && $-0.14 \pm 0.03$  & $-0.14 \pm 0.03$  \\
    \bottomrule
    \medskip
    \end{tabular*}
\end{table*}
\endgroup

\subsection{Combining the spectra}
\label{sec:results_comparison}

In this section, we compare the differences between the measured sodium line depths when the spectra have been combined using the weighted and simple (unweighted) averages -- all results discussed are displayed in Table~\ref{tab:all_results}. Two sets of results have been included for MASCARA-2b and WASP-76b to make comparisons with previously published work (see section~\ref{sec:results_previous_work}) which have used different nights of data.

First, the measured depths using both approaches for WASP-69b, WASP-21b, WASP-79b, WASP-76b (three nights), WASP-76b (five nights), MASCARA-2b (one night), MASCARA-2b (three nights), WASP-121b, and KELT-9b all agree within 1$\sigma$ uncertainties, and we conclude that either method is sufficient for these targets.

For WASP-189b, the errors are much lower than other targets due to the brightness of the host star and the availability of four nights of data. The measured line depths using both methods are consistent overall, but the agreement between the D2 line depths is at the extremity of the 1$\sigma$ uncertainty range. This could possibly arise from imperfect removal of ISM sodium contamination, although the results did not change in an independent test when the Doppler shift for stellar reflex motion was ignored (see section~\ref{sec:method_velocity}).

Two targets remain for more thorough discussion: HD~189733~b and WASP-49b. Using the weighted combination, we measured absorption depths which were more than 1$\sigma$ lower than the unweighted method. Both methods confirm the detection of Na in the atmospheres of these planets, however there is slight disagreement in the strength of the lines. The D1 lines for HD~189733~b and the D2 lines for WASP-49b agree within 1$\sigma$ uncertainties, but the other doublet lines for each planet vary more significantly. Several factors could lead to these differences. First, the number of in-transit observations for each night is relatively low compared to the other planets in this survey. This affects the SNR of the combined transmission spectrum. Next, the location of the planetary signal in the 2D map of residual spectra lies mostly inside the low-SNR region produced by the deep stellar sodium line cores. Therefore, the unweighted results may be falsely enhanced by low-SNR residual features which do not originate from the planet, and the weighted approach may give a better representation of the true planetary signal. Finally, the location of the planetary absorption also overlaps with the majority of the features of the CLV and RM model, which may add further uncertainty. Some of these factors are also evident in other planets from this survey, although not all at once. It is therefore most likely a combination of these reasons that causes the difference in the measured line depths.

Over the last few years, the issues caused by low flux in the cores of deep stellar lines have become a well-known problem and the weighted average approach is now commonly employed for high-resolution transmission spectroscopy \citep[e.g.][]{allart2020, chen2020b, borsa2021, sedaghati2021}. Since there is a difference between the weighted and unweighted results for two planets in this survey (HD~189733~b and WASP-49b), performing a homogeneous analysis is important when comparing the extent of sodium absorption across the sample of exoplanets. The following discussions in section~\ref{sec:trends} refer to the weighted results only, however we continue to show the unweighted results in the tables for completeness and comparison to other work.

\subsection{Comparison with previous work}
\label{sec:results_previous_work}

Many of the archival datasets used within this work have previously been analysed by several authors -- these results are also shown in Table~\ref{tab:all_results}. For comparison and verification, we reproduced the relevant figures from the listed references (by presenting our results in the same style), which can be found online at \href{https://osf.io/g3z6r/}{osf.io/g3z6r}. All of the weighted and unweighted results agree with the literature values within 1$\sigma$ uncertainties, except for HD~189733~b, WASP-69b, and WASP-76b (three nights).

Using the unweighted approach, the measured Na line depths for HD~189733~b mostly agree with the literature values within 1$\sigma$, except for the D2 line when compared to \citet{casasayas-barris2017}. However, the weight-combined results are all shallower than the same measurements made by \citet{wyttenbach2015}, \citet{casasayas-barris2017}, and \citet{langeveld2021}. We note that in \citet{langeveld2021}, the spectra are only weighted when combining the nights, unlike the approach to weight each individual spectrum in this work.
For WASP-69b, our measured D2 line depths are shallower than the value from \citet{casasayas-barris2017} by more than 2$\sigma$. However, the depths closely agree with the results presented by \citet{khalafinejad2021}, who report a D2 line depth of $-3.2~\pm~0.3$~\% and a D1 line depth of $-1.2~\pm~0.3$~\% using one night of data from CARMENES.
For WASP-76b (three nights), our results are consistent with \citet{zak2019}, but the weight-combined D2 line depth of $-0.52 \pm 0.06$~\% is deeper by more than 1$\sigma$ than the $-0.33 \pm 0.09$~\% measurement presented by \citet{seidel2019}.

We also note that the weight-combined results for WASP-49b are shallower than the same measurements made by \citet{wyttenbach2017}, however the uncertainties are large due to the low SNR of the stellar spectra and only having access to a low number of in-transit exposures. There is agreement between the \mbox{WASP-49b} unweighted results and the referenced values, which may be explained if \citet{wyttenbach2017} used the unweighted approach. 

These discrepancies may arise due to various factors, including: (1) combining the spectra using weighted or simple averages; (2) differing telluric correction methods; (3) analysing the one-dimensional (s1d) spectra instead of individual orders of the two-dimensional (e2ds) spectra; (4) exclusion of observations during the transit ingress or egress; or (5) variations in the data analysis pipelines, RM/CLV models, and Gaussian fitting algorithms. All of these factors vary between studies and can impact the resultant transmission spectrum, highlighting the importance of conducting a homogeneous survey.

\section{Sodium trends in gas giant atmospheres}
\label{sec:trends}
With sodium being a commonly detected species in gas giant atmospheres, it is valuable to understand how the Na line absorption relates to the bulk properties of the planet and its host star. In particular, we would like to understand the physical processes that take place in the layer of atmosphere where sodium is present, and use this information to make predictions for the general population of planets. In this section we present our results and search for trends relating to atmospheric heights, line ratios, and wind speeds.

\subsection{Relative height of the sodium D2 line}
\label{sec:trends_heights}

The sodium D2 line is clearly resolved and measured for all planets in this survey, whereas the D1 depths are less significant in some cases (e.g. WASP-69b and WASP-189b). We therefore only used the measured D2 depths in the following analysis, however we repeated the process using the D1 depths and were able to make similar conclusions (only with more uncertainty due to the less significant measurements for some planets).

The depth of the absorption lines can be related to an atmospheric height at which sodium is present using the following derivation. First, the scale height $\left(H_{\text{sc}}\right)$ of the atmosphere is defined as: 
\begin{equation}
    H_{\text{sc}} = \frac{k_{\text{B}}T_{\text{eq}}}{\mu_{\text{m}}g} ~,
    \label{eqn:scale_height}
\end{equation}
where $k_{\text{B}}$ is the Boltzmann constant, $T_{\text{eq}}$ is the equilibrium temperature, $\mu_{\text{m}}$ is the mean molecular weight, and $g$ is the planet's surface gravity. For a hot Jupiter with a H/He-dominated, solar composition atmosphere, we assume $\mu_{\text{m}} = 2.4$~u. The observable atmosphere can typically extend around 5--10 scale heights \citep{madhusudhan2019}.

Next, an opaque transiting planet with no atmosphere will have a white-light transit depth equal to the ratio of the sky-projected areas of the planet and star:
\begin{equation}
    \Delta_0 = \left(\frac{R_{\text{p}}}{R_\ast}\right)^2 ~.
    \label{eqn:white_light_transit}
\end{equation}
If the planet has an atmosphere, molecular or atomic absorption of light from the star will cause the atmosphere to appear opaque when viewed at certain wavelengths, which increases the apparent radius. Therefore, a wavelength-dependent transit depth can similarly be defined as 
\begin{equation}
    \Delta_\lambda = \left(\frac{R_{\text{p}} + H_\lambda}{R_\ast}\right)^2 = \left(\frac{R_{\text{p}}}{R_\ast}\right)^2 + \frac{2R_{\text{p}}H_\lambda}{R_\ast^2} + \left(\frac{H_\lambda}{R_\ast}\right)^2 ~,
    \label{eqn:wavelength_transit}
\end{equation}
where $H_\lambda$ is the additional height of the atmosphere where the absorbing species is present. Combining equations~\ref{eqn:white_light_transit} and \ref{eqn:wavelength_transit} gives 
\begin{equation}
    \Delta_\lambda = \Delta_0 + 2\Delta_0\frac{H_\lambda}{R_{\text{p}}} + \Delta_0\left(\frac{H_\lambda}{R_{\text{p}}}\right)^2 ~.
    \label{eqn:wavelength_transit_2}
\end{equation}
The amount of absorption at a particular wavelength ($\delta_\lambda$) is the difference between the wavelength-dependent and white-light transit depths, expressed as 
\begin{equation}
    \delta_\lambda = \Delta_\lambda - \Delta_0 = 2\Delta_0\frac{H_\lambda}{R_{\text{p}}} + \Delta_0\left(\frac{H_\lambda}{R_{\text{p}}}\right)^2 ~.
    \label{eqn:absorption_depth}
\end{equation}
This is related to the calculated transmission spectrum from equation~\ref{eqn:transmission_combined}, where
\begin{equation}
    \delta_\lambda = 1 - \frac{F_{\lambda, \text{in}}}{F_{\lambda, \text{out}}} = -{\Re'}_{\lambda} ~.
    \label{eqn:absorption_depth_Re}
\end{equation}
When comparing absorption from a sample of planets, it is difficult to justify looking only at the measured line depths since each planet has a different white-light transit depth (equation~\ref{eqn:white_light_transit}). A more useful quantity to evaluate is the ratio of the height of the atmosphere to the planetary radius, hereafter denoted as $h_{\lambda} = H_{\lambda}/R_{\text{p}}$. 
Rearranging equation~\ref{eqn:absorption_depth} gives 
\begin{equation}
    \frac{\delta_\lambda}{\Delta_0} = 2h_{\lambda} + {h_{\lambda}}^2 ~.
    \label{eqn:absorption_depth_quadratic}
\end{equation}
The solution to this quadratic equation therefore allows for the calculation of the relative height: 
\begin{equation}
    h_{\lambda} = \sqrt{1 + \frac{\delta_\lambda}{\Delta_0}} - 1 ~.
    \label{eqn:height_radius_ratio}
\end{equation}
This quantity was calculated using the measured sodium D2 line depths to compare the relative heights of the sodium layer ($h_{\text{Na}}$) for each planet. The negative quadratic solution is ignored due to the non-physical interpretation.

\begingroup
\renewcommand{\arraystretch}{1.15} 
\begin{table}
    \centering
    \caption{Equilibrium temperature ($T_{\text{eq}}$), surface gravity ($g$), atmospheric scale height $\left(H_{\text{sc}}\right)$, and $\xi$ value for the ten gas giant planets analysed in this survey. Surface gravity and scale heights were derived using the parameters listed in Tables~\ref{tab:parameters1-5} and \ref{tab:parameters6-10}.}
    \label{tab:scale_heights}
    \begin{tabular*}{\columnwidth}{l@{\extracolsep{\fill}}lllll}
    \toprule
    Planet          && $T_{\text{eq}}$ (K)  & $g$ (m s$^{-2}$)      & $H_{\text{sc}}$ (km)  & \multicolumn{1}{c}{$\xi$} \\
    \midrule
    WASP-69b        && $960\pm20$           & $6.0\pm0.5$           & $550\pm40$            & $0.23\pm0.02$ \\
    HD~189733~b     && $1200\pm15$          & $22.7\pm0.9$          & $180\pm10$            & $1.10\pm0.05$ \\
    WASP-21b        && $1330\pm40$          & $5.3\pm0.5$           & $870\pm90$            & $0.28\pm0.03$ \\
    WASP-49b        && $1450\pm40$          & $7.2\pm0.7$           & $700\pm70$            & $0.42\pm0.04$ \\
    WASP-79b        && $1720\pm30$          & $9.4\pm1.0$           & $630\pm60$            & $0.65\pm0.07$ \\
    WASP-76b        && $2210\pm30$          & $6.7\pm0.4$           & $1140\pm70$           & $0.60\pm0.04$ \\
    MASCARA-2b      && $2350\pm50$          & $27.2\pm1.5$          & $300\pm20$            & $2.58\pm0.15$ \\
    WASP-121b       && $2360\pm60$          & $8.8\pm0.6$           & $930\pm60$            & $0.84\pm0.06$ \\
    WASP-189b       && $2640\pm40$          & $19.7\pm1.6$          & $460\pm40$            & $2.10\pm0.18$ \\
    KELT-9b         && $3670\pm150$         & $19.9\pm5.7$          & $640\pm190$           & $2.94\pm0.85$ \\
    \bottomrule
    \end{tabular*}
\end{table}
\endgroup

We performed a search for trends between the measured relative heights and the planetary and stellar parameters. In general, the hotter and more massive planets had lower relative heights ($\sim0.12$), and the colder and lower mass planets had larger relative heights ($\gtrsim0.3$). 
We deduce that this is due to two fundamental properties of the planet: equilibrium temperature (which relates to the incident stellar irradiation) and surface gravity (which relates to the planetary mass and radius). 
Since these two factors are independent and do not influence each other, they should both be considered simultaneously when looking for trends.

\begin{figure*}
    \centering
    \includegraphics[width=0.98\textwidth]{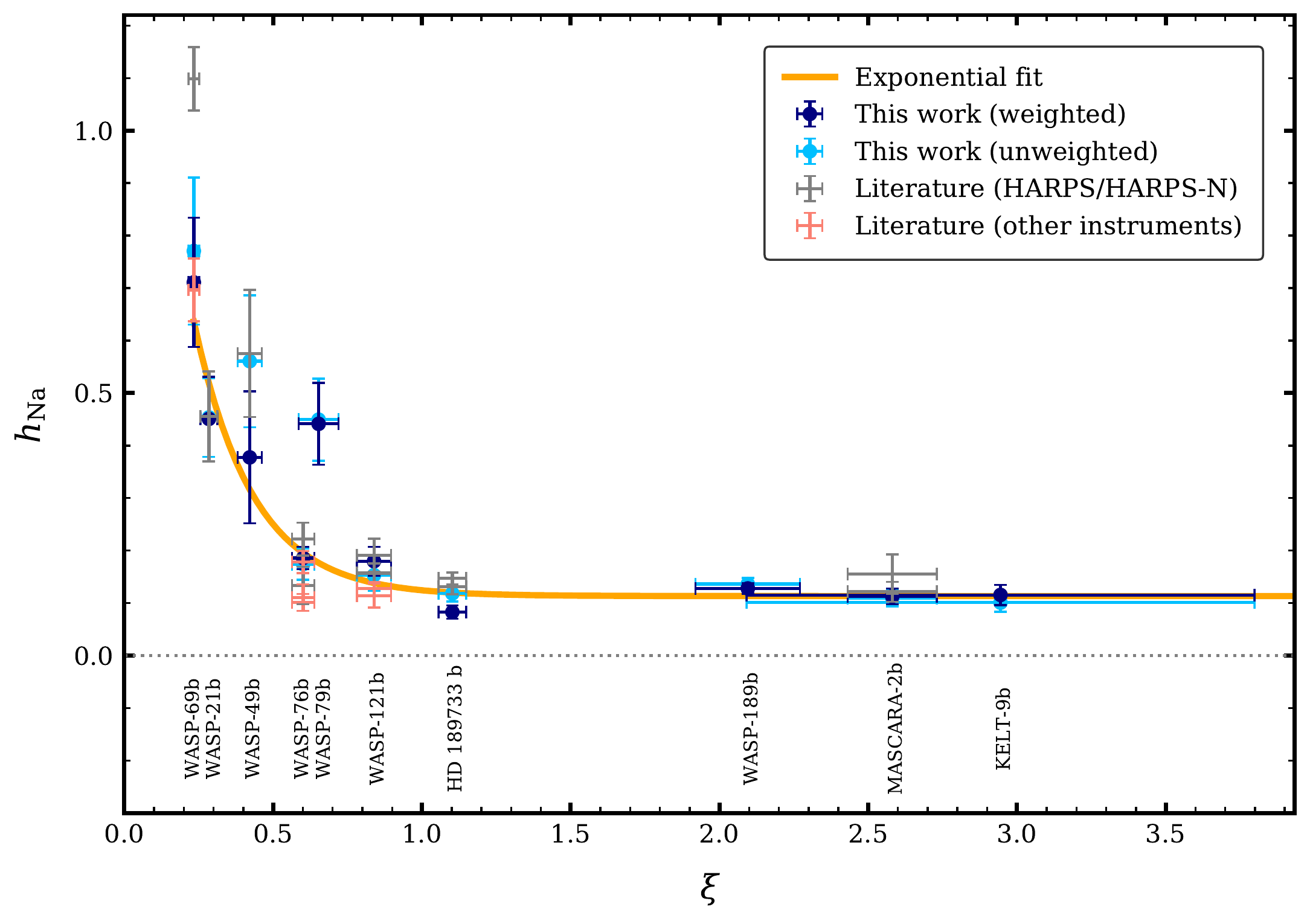}
    \caption{Relationship between the relative height of sodium in the ten gas giant atmospheres and $\xi$. The relative height, $h_{\text{Na}}$, is calculated using the measured sodium D2 line depths for the weight-combined spectra (dark-blue), unweighted spectra (light-blue), literature values using the same HARPS/HARPS-N data (grey), and literature values using other high-resolution spectrographs (salmon). References for results from other instruments: WASP-69b with CARMENES \citep{khalafinejad2021}, WASP-76b with ESPRESSO \citep{tabernero2021} and GRACES \citep{deibert2021}, and WASP-121b with ESPRESSO \citep{borsa2021}. The exponential curve described by equation~\ref{eqn:height_trend} is shown in orange and was fitted to the weight-combined results.
    \smallskip}
    \label{fig:height_ratio_trend}
\end{figure*}

\begin{figure}
    \centering
    \includegraphics[width=\columnwidth]{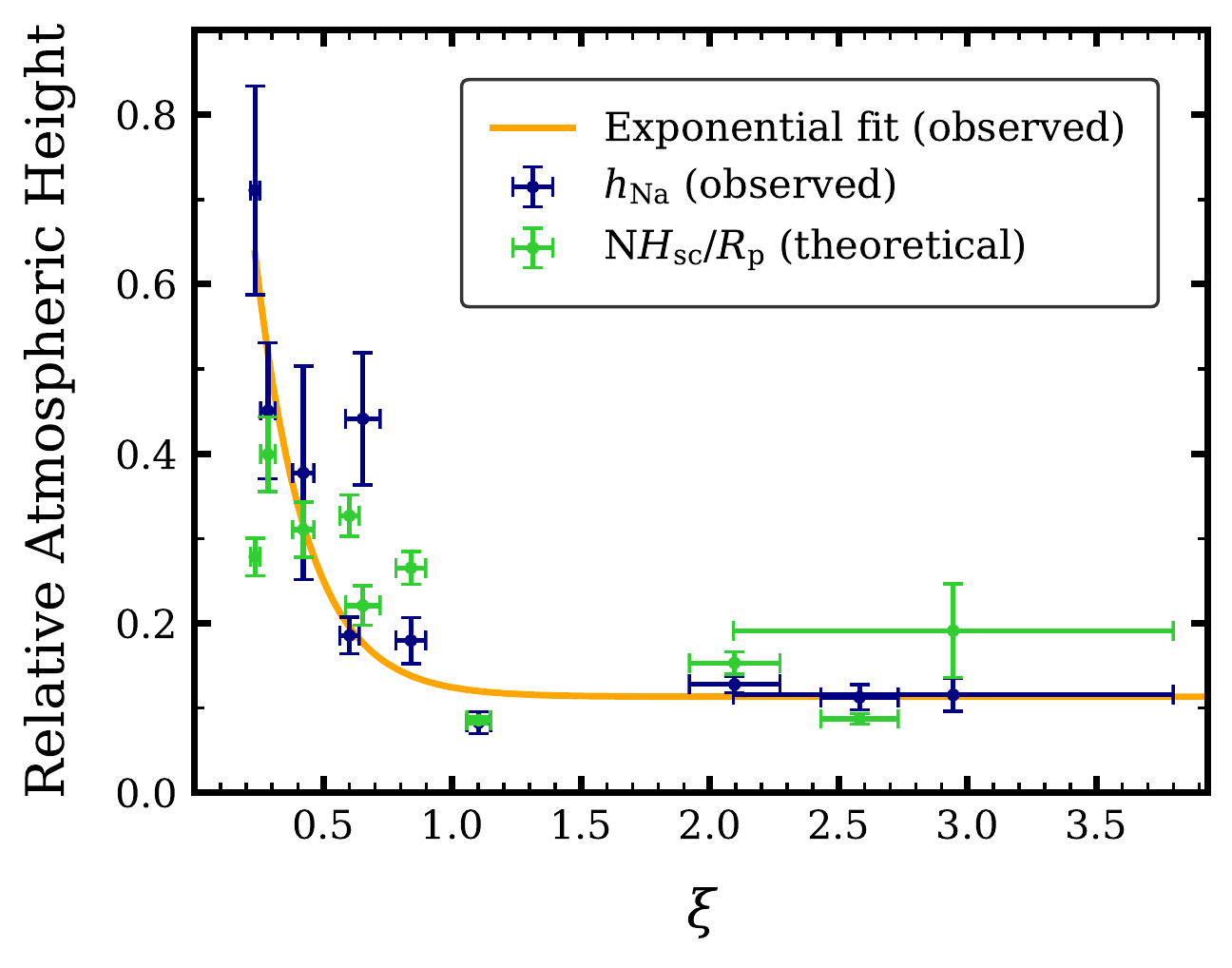}
    \vspace{-2em}
    \caption{Observed relative atmospheric heights ($h_{\text{Na}}$) from Figure~\ref{fig:height_ratio_trend} (dark-blue) together with the best-fitting exponential (orange). The theoretical atmospheric scale heights normalised to the planetary radii and multiplied by a constant (N$H_{\text{sc}}/R_{\text{p}}$) are shown in green for comparison. The constant N is chosen such that the observed and theoretical heights are identical for the lowest $h_{\text{Na}}$ (for HD~189733~b, with $\xi \sim 1.1$).}
    \label{fig:height_ratio_and_scale_heights}
\end{figure}

We introduce the quantity $\xi$ to represent a scaled product of equilibrium temperature ($T_{\text{eq}}$) and surface gravity ($g$):
\begin{equation}
    \xi = \left(\frac{T_{\text{eq}}}{1000~\text{K}}\right)\left(\frac{g}{g_J}\right) ~,
    \label{eqn:xi}
\end{equation}
where $g_{\text{J}}$ is the surface gravity of Jupiter. The calculated $\xi$ values are listed in Table~\ref{tab:scale_heights}, along with the equilibrium temperature, surface gravity, and scale height for each planet. Figure~\ref{fig:height_ratio_trend} shows the relative height of sodium for all ten planets against $\xi$. The weight-combined results are shown in dark-blue, together with the unweighted results (light-blue) and literature values using the same HARPS/HARPS-N data (grey). The salmon-coloured points show recent results from other high-resolution spectrographs: WASP-69b with CARMENES \citep{khalafinejad2021}, WASP-76b with ESPRESSO \citep{tabernero2021} and GRACES \citep{deibert2021}, and WASP-121b with ESPRESSO \citep{borsa2021}. We find that $h_{\text{Na}}$ is well described by an exponential trend of the form 
\begin{equation}
    h_{\lambda} = a\text{e}^{-b\xi} + c ~.
    \label{eqn:height_trend}
\end{equation}
As shown in Figure~\ref{fig:height_ratio_trend}, we fit this curve to the weight-combined results using the \texttt{optimize.curve\_fit} function from \texttt{scipy}. 
The best-fitting values for the variables are: 
$a=1.70 \pm 1.04$, 
$b=5.04 \pm 1.63$, and 
$c=0.113 \pm 0.013$. 
The reduced chi-square of the fit across the full sample is $\chi^2_\nu = 3.7$, much of which is skewed by two planets that deviate from the fit by more than 3$\sigma$ (WASP-79b and HD~189733~b). The reduced chi-square when excluding these two planets is $\chi^2_\nu = 1.2$.
Since the exponential curve is asymptotic to the value of $c$, our results suggest that planets with $\xi\gtrsim1.25$ are likely to have an upper limit on $h_{\text{Na}}$ of $\sim0.113$. The sodium features could potentially be muted due to other atmospheric effects, such as high-altitude clouds and hazes in lower temperature planets, or ionisation of most of the sodium in extremely irradiated environments. Therefore, results lower than those given by equation~\ref{eqn:height_trend} are also possible. 
The underlying physical processes behind this trend are not discussed in this current work, but these results motivate further observations to confirm the trend and theoretical studies to investigate possible physical mechanisms. 
Further refinement of this curve will be possible with more observations of low-$\xi$ planets.

All the planets in our sample seem to generally follow this trend except for WASP-79b, which shows a significantly higher atmospheric height. Since we only have access to one observed transit of this planet with few out-of-transit frames, there are large uncertainties and the continuum of the combined transmission spectrum is noisier in comparison to several other targets. WASP-79b is therefore a good candidate for follow-up observations, and combining more transits would help to refine the system parameters, line depths, and atmospheric height. We also note that for HD~189733~b, our weighted and unweighted results for the D2 line are different by more than 1$\sigma$. The weighted result is significantly lower than the trend, although the unweighted and literature values agree with it. Some possible reasons for this discrepancy are discussed in section~\ref{sec:results_comparison}.

Theoretically, in ideal conditions where observations probe the same number of scale heights for all planets and the observable terminator is isothermal with a temperature $T_{\text{eq}}$ (or a constant fraction of it), one could assume that the probed atmospheric height is proportional to a uniform scale height (equation~\ref{eqn:scale_height}). Such an assumption is unlikely to be realistic across the diverse sample of planets considered here. Nevertheless, calculating a normalised scale height for each planet, N$H_{\text{sc}}/R_{\text{p}}$ (i.e. scale height normalised by the planetary radius and multiplied by a factor N), allows for direct comparison of the observed heights to the idealised theoretical expectations. This comparison is shown in Figure~\ref{fig:height_ratio_and_scale_heights}, where the weighted results (dark-blue) and exponential trend (orange) from Figure~\ref{fig:height_ratio_trend} are plotted alongside the theoretical normalised scale heights (green). The value of N was chosen such that the theoretical and observed values are identical for the planet HD~189733~b (with $\xi \sim 1.1$). The theoretical values show a comparable trend to some of the observed values. However, there are significant deviations in several of the cases. These differences may suggest departures from idealised conditions, such as the sodium detections probing a different number of scale heights for each planet, or that $T_{\text{eq}}$ is not a good approximation for the temperatures at these altitudes. The physical mechanisms responsible for possible variations in temperature structures warrant further investigation, but could be due to thermospheric absorption of high-energy stellar radiation \citep[e.g.][]{koskinen2013, wyttenbach2017, wyttenbach2020} or thermal inversions \citep[e.g.][]{fortney2008, madhusudhan2010, yan2020, yan2022}.
The observed transit depths reported here provide new impetus for further observations and theoretical work to investigate possible explanations of the trend and deviations thereof.

\subsection{Predicting sodium absorption for other planets}

\begingroup
\renewcommand{\arraystretch}{1.45} 
\begin{table*}
    \centering
    \caption{Parameters adopted in this work to estimate the extent of sodium in the atmospheres of five highly irradiated giant exoplanets. 
        The quantities displayed are: V-band apparent magnitude ($m_{\text{V}}$), stellar effective temperature ($T_{\text{eff}}$), stellar radius ($R_\ast$), semi-major axis ($a$), planetary mass ($M_{\text{p}}$), planetary radius ($R_{\text{p}}$), equilibrium temperature ($T_{\text{eq}}$), planetary surface gravity ($g$), scaled product of equilibrium temperature and surface gravity ($\xi$), estimated relative atmospheric height of sodium ($h_{\text{Na}}$), and estimated sodium D2 line depth ($\delta_{\text{Na-D2}}$). Planets are ordered by increasing $\xi$. 
        ${^\dagger}$~Derived values. 
        ${^\ddagger}$~Estimates when adopting the central values of the parameters used in equations~\ref{eqn:height_radius_ratio} and \ref{eqn:height_trend}. 
        References to listed parameters: 
        (A)~\citet{yu2018}, 
        (B)~\citet{bakos2021}, 
        (C)~\citet{eastman2016}, 
        (D)~\citet{zhou2016}, 
        (E)~\citet{queloz2010}.
        }
    \label{tab:case_study}
    \begin{tabular*}{\textwidth}{l@{\extracolsep{\fill}}llllll}
    \toprule
    Parameter           & Unit              & K2-232~b                                          & HAT-P-60~b                                        & KELT-4~A~b                                         & KELT-17b                                          & WASP-8b                                             \\
    \midrule
    $m_{\text{V}}$      & mag.              & 9.82                                              & 9.72                                              & 9.98                                               & 9.23                                              & 9.79                                                \\
    $T_{\text{eff}}$    & K                 & 5888 $^{+71}_{-77}$ $^{(\text{A})}$               & 6212 $^{+26}_{-26}$ $^{(\text{B})}$               & 6206 $^{+75}_{-75}$ $^{(\text{C})}$                & 7454 $^{+49}_{-49}$ $^{(\text{D})}$               & 5600 $^{+80}_{-80}$ $^{(\text{E})}$                 \\
    $R_\ast$            & $R_{\sun}$        & 1.233 $^{+0.026}_{-0.025}$ $^{(\text{A})}$        & 2.197 $^{+0.027}_{-0.020}$ $^{(\text{B})}$        & 1.603 $^{+0.039}_{-0.038}$ $^{(\text{C})}$         & 1.645 $^{+0.060}_{-0.055}$ $^{(\text{D})}$        & 0.945 $^{+0.051}_{-0.051}$ $^{(\text{E})}$          \\
    $a$                 & A.U.              & 0.0991 $^{+0.0022}_{-0.0021}$ $^{(\text{A})}$     & 0.06277 $^{+0.00017}_{-0.00017}$ $^{(\text{B})}$  & 0.04317 $^{+0.00079}_{-0.00074}$ $^{(\text{C})}$   & 0.04881 $^{+0.00065}_{-0.00061}$ $^{(\text{D})}$  & 0.0801 $^{+0.0014}_{-0.0016}$ $^{(\text{E})}$       \\
    $M_{\text{p}}$      & $M_{\text{J}}$    & 0.387 $^{+0.044}_{-0.042}$ $^{(\text{A})}$        & 0.574 $^{+0.038}_{-0.038}$ $^{(\text{B})}$        & 0.902 $^{+0.060}_{-0.059}$ $^{(\text{C})}$         & 1.31 $^{+0.28}_{-0.29}$ $^{(\text{D})}$           & 2.244 $^{+0.079}_{-0.093}$ $^{(\text{E})}$          \\
    $R_{\text{p}}$      & $R_{\text{J}}$    & 1.058 $^{+0.023}_{-0.022}$ $^{(\text{A})}$        & 1.631 $^{+0.024}_{-0.024}$ $^{(\text{B})}$        & 1.699 $^{+0.046}_{-0.045}$ $^{(\text{C})}$         & 1.525 $^{+0.065}_{-0.060}$ $^{(\text{D})}$        & 1.038 $^{+0.007}_{-0.047}$ $^{(\text{E})}$          \\
    $T_{\text{eq}}$     & K                 & 1000 $^{+20}_{-20}$ $^{\dagger}$                  & 1770 $^{+10}_{-10}$ $^{\dagger}$                  & 1820 $^{+30}_{-30}$ $^{\dagger}$                   & 2090 $^{+40}_{-40}$ $^{\dagger}$                  & 930 $^{+30}_{-30}$ $^{\dagger}$                     \\
    $g$                 & m s$^{-2}$        & 8.9 $^{+1.1}_{-1.0}$ $^{\dagger}$                 & 5.6 $^{+0.4}_{-0.4}$ $^{\dagger}$                 & 8.1 $^{+0.6}_{-0.6}$ $^{\dagger}$                  & 14.6 $^{+3.2}_{-3.3}$ $^{\dagger}$                & 54.0 $^{+2.0}_{-4.1}$ $^{\dagger}$                  \\
    $\xi$               &                   & 0.36 $^{+0.04}_{-0.04}$ $^{\dagger}$              & 0.40 $^{+0.03}_{-0.03}$ $^{\dagger}$              & 0.59 $^{+0.05}_{-0.05}$ $^{\dagger}$               & 1.23 $^{+0.27}_{-0.28}$ $^{\dagger}$              & 2.03 $^{+0.10}_{-0.17}$ $^{\dagger}$                \\
    $h_{\text{Na}}$     &                   & $0.388$ $^{\ddagger}$                             & $0.340$ $^{\ddagger}$                             & $0.198$ $^{\ddagger}$                              & $0.116$ $^{\ddagger}$                             & $0.113$ $^{\ddagger}$                                \\
    $\delta_{\text{Na-D2}}$ & \%            & $0.69$ $^{\ddagger}$                              & $0.44$ $^{\ddagger}$                              & $0.49$ $^{\ddagger}$                               & $0.21$ $^{\ddagger}$                              & $0.29$ $^{\ddagger}$                                 \\
    \bottomrule
    \end{tabular*}
\end{table*}
\endgroup

The suggested trend between atmospheric height and $\xi$ can be used to predict which planets may or may not show significant sodium absorption. This could help with optimally planning observations and selecting good targets for high-resolution transmission spectroscopy. As an example, we chose five highly irradiated giant exoplanets with a range of equilibrium temperatures and surface gravity, and assessed their potential for the detection of sodium absorption. These planets and their relevant parameters are listed in Table~\ref{tab:case_study}, ordered by increasing $\xi$. We used equations~\ref{eqn:height_radius_ratio} and \ref{eqn:height_trend} to estimate the relative atmospheric height of sodium ($h_{\text{Na}}$) and the D2 line depth ($\delta_{\text{Na-D2}}$). All systems have V-band apparent magnitudes $<10$ which is conducive for observing with 4~m class telescopes and instruments such as HARPS.

The first three planets in Table~\ref{tab:case_study} (K2-232~b, HAT-P-60~b, and \mbox{KELT-4~A~b}) all have $\xi<1$. Under the assumption that these planets follow the trend indicated in Figure~\ref{fig:height_ratio_trend}, their estimated D2 line depths are 0.69~\%, 0.44~\%, and 0.49~\% respectively. These depths are large enough for Na detections to be verifiable with future HARPS/HARPS-N observations of two or three transits per planet. In contrast, the predicted D2 line depths for KELT-17b (0.21~\%) and WASP-8b (0.29~\%) are lower due to their higher $\xi$ values. 
These depths are comparable despite their significantly different equilibrium temperatures and surface gravity, reinforcing the importance of considering both of these properties simultaneously. We were only able to measure lines this shallow for the brightest targets in this survey (e.g. those with $m_{\text{V}}<8$). 
It is therefore unlikely that the Na doublet would be detectable for KELT-17b ($m_{\text{V}} = 9.23$) and \mbox{WASP-8b} ($m_{\text{V}} = 9.79$) using a HARPS-like facility. This is consistent with recent results, as \citet{wyttenbach2017} reported a flat WASP-8b spectrum with an upper limit for sodium absorption of $<0.54$~\%, and \citet{stangret2021} were unable to detect sodium in KELT-17b.

Deviations from this trend are possible, which may inform us about the diversity of atmospheric processes among giant exoplanets and motivate new theoretical studies in this direction. Therefore, the trend should not be used as the only method for assessing detectable Na features, but as a complimentary evaluation in addition to other observational factors. We note that our sample includes only planets for which HARPS/\mbox{HARPS-N} observations show detectable Na features. It is possible that some planets, particularly those with low-$\xi$ values, may not show significant Na features due to the presence of high-altitude clouds/hazes \citep[e.g.][]{allart2020, chen2020a} and hence significantly deviate from the trend predicted here.

\begin{figure*}
    \centering
    \includegraphics[width=0.98\textwidth]{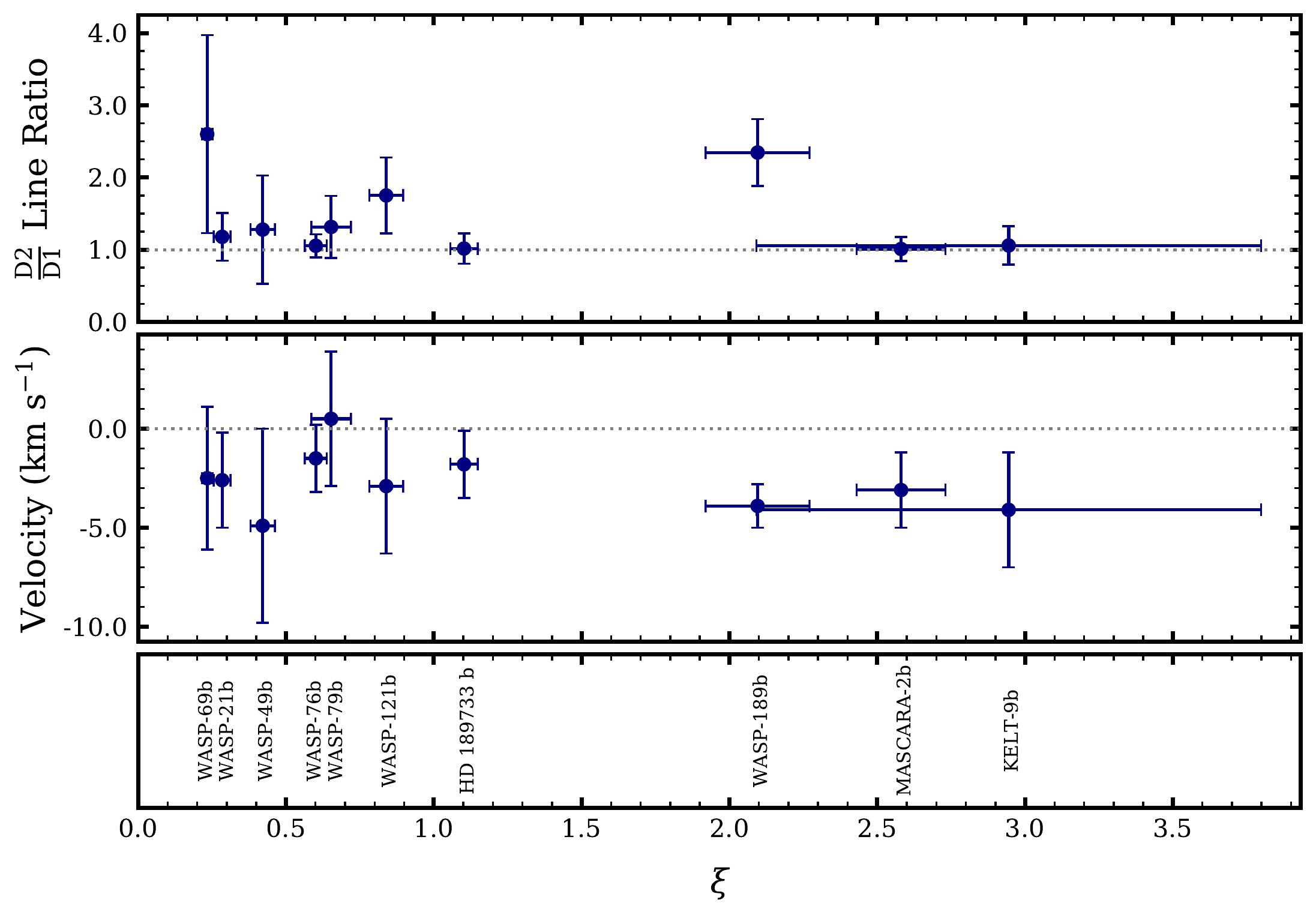}
    \caption{Measured sodium D2/D1 line ratios (top panel) and velocity offsets of the Gaussian centroids (second panel) for all ten planets, against their $\xi$ value. The horizontal dotted line in the top panel indicates a line ratio of 1 -- seven planets are consistent with this (within 1$\sigma$ uncertainties), which could suggest similar atmospheric structures. The second panel shows that this sample of ten irradiated gas giants exhibit consistent blueshifted velocities (below the horizontal dotted line), indicating the presence of net day-night atmospheric winds in a diverse range of giant exoplanets. All values are calculated using the weight-combined spectra, and further discussion can be found in sections~\ref{sec:ratios} and \ref{sec:winds}.
    \medskip\medskip\medskip}
    \label{fig:line_ratios_winds}
\end{figure*}

\begingroup
\renewcommand{\arraystretch}{1.25} 
\begin{table*}
    \centering
    \caption{Measured sodium D-line absorption depths and corresponding D2/D1 line ratios for the ten gas giant planets in this survey, using all available nights of data listed in Table~\ref{tab:observing_log}. The second, third, and fourth columns are the results for the weight-combined approach, and the fifth, sixth, and seventh columns are the same results for the unweighted approach.
    \medskip
    }
    \label{tab:final_results}
    \begin{tabular*}{\textwidth}{l@{\extracolsep{\fill}}lllclllc}
    \toprule
    \multirow{2}{*}{Planet} && \multicolumn{3}{c}{Weight-combined spectra}                      && \multicolumn{3}{c}{Unweighted spectra} \\
        && \multicolumn{1}{c}{D2~(\%)} & \multicolumn{1}{c}{D1~(\%)}    & D2/D1 Ratio           && \multicolumn{1}{c}{D2~(\%)} & \multicolumn{1}{c}{D1~(\%)}    & D2/D1 Ratio   \\
    \midrule
    WASP-69b                && $-3.28 \pm 0.70$  & $-1.26 \pm 0.61$     & $2.60 \pm 1.37$       && $-3.64 \pm 0.82$  & $-1.19 \pm 0.80$     & $3.05 \pm 2.16$                   \\
    HD~189733~b             && $-0.39 \pm 0.06$  & $-0.39 \pm 0.06$     & $1.02 \pm 0.21$       && $-0.57 \pm 0.07$  & $-0.47 \pm 0.07$     & $1.22 \pm 0.24$                   \\
    WASP-21b                && $-1.16 \pm 0.22$  & $-0.99 \pm 0.20$     & $1.18 \pm 0.33$       && $-1.17 \pm 0.20$  & $-0.96 \pm 0.19$     & $1.22 \pm 0.32$                   \\
    WASP-49b                && $-1.20 \pm 0.46$  & $-0.94 \pm 0.42$     & $1.28 \pm 0.75$       && $-1.93 \pm 0.51$  & $-2.01 \pm 0.47$     & $0.96 \pm 0.34$                   \\
    WASP-79b                && $-1.12 \pm 0.23$  & $-0.85 \pm 0.22$     & $1.31 \pm 0.43$       && $-1.14 \pm 0.23$  & $-0.72 \pm 0.23$     & $1.58 \pm 0.59$                   \\
    WASP-76b                && $-0.47 \pm 0.05$  & $-0.45 \pm 0.05$     & $1.05 \pm 0.16$       && $-0.44 \pm 0.08$  & $-0.49 \pm 0.07$     & $0.89 \pm 0.20$                   \\
    MASCARA-2b              && $-0.31 \pm 0.04$  & $-0.31 \pm 0.04$     & $1.00 \pm 0.17$       && $-0.31 \pm 0.04$  & $-0.31 \pm 0.04$     & $0.99 \pm 0.19$                   \\
    WASP-121b               && $-0.65 \pm 0.10$  & $-0.37 \pm 0.09$     & $1.75 \pm 0.52$       && $-0.54 \pm 0.11$  & $-0.28 \pm 0.09$     & $1.93 \pm 0.75$                   \\
    WASP-189b               && $-0.13 \pm 0.01$  & $-0.05 \pm 0.01$     & $2.35 \pm 0.46$       && $-0.14 \pm 0.01$  & $-0.07 \pm 0.01$     & $1.89 \pm 0.37$                   \\
    KELT-9b                 && $-0.16 \pm 0.03$  & $-0.16 \pm 0.03$     & $1.06 \pm 0.27$       && $-0.14 \pm 0.03$  & $-0.14 \pm 0.03$     & $1.05 \pm 0.28$                   \\
    \bottomrule
    \end{tabular*}
\end{table*}
\endgroup

\subsection{Line ratios}
\label{sec:ratios}
Previous studies have investigated how the D2/D1 line ratio can be used to learn about atmospheric and exospheric structure (e.g. hydrostatic or non-hydrostatic), and aid current understanding of properties such as: variability of chemical composition and mixing ratios \citep{slanger2005}, the velocity distribution of the Na atoms, and in which regime (e.g. optically thick or thin) the absorption occurs \citep{oza2019, gebek2020}. The measured sodium absorption depths for the ten planets in this survey (using both the weighted and unweighted approaches) are summarised in Table~\ref{tab:final_results}, together with the corresponding D2/D1 line ratios. These results may provide a good basis for further research in this area.

In addition to searching for trends in the relative height of atmospheric sodium, we also attempted to see if there are any similarities between the line ratios. The top panel of Figure~\ref{fig:line_ratios_winds} shows the measured D2/D1 ratio for the weight-combined spectra, against the $\xi$ quantity discussed in section~\ref{sec:trends_heights}. No clear trend can be seen, although seven of the planets have a ratio that is consistent with 1 (within uncertainties). Three planets (WASP-69b, WASP-121b, and WASP-189b) all have line ratios greater than 1, which could suggest that these planets share common atmospheric structures despite their different planetary properties \citep{gebek2020}.

The atmosphere of WASP-121b has recently been studied by \citet{borsa2021} using ESPRESSO: one transit was observed with the 1-UT HR21 mode and a second transit with the 4-UT MR42 mode. The measured sodium D2 line was comparable within 1$\sigma$ for both observing modes, however the D1 line varied more significantly. This meant that two different line ratios were measured: \mbox{$1.47\pm0.47$} for the 1-UT mode, and $0.88\pm0.12$ for the 4-UT mode. Our HARPS-measured D2/D1 ratio of $1.75\pm0.52$ is only consistent with the 1-UT result. Since different instruments and resolutions yield different results, it is possible that line ratios (and therefore line depths) may be impacted by external factors such as instrumental systematics, resolving power, and weather effects. The calculated D2/D1 line ratios using the unweighted results (listed in Table~\ref{tab:final_results}) are more varied among the sample of planets, which also suggests that differences could arise from the methods of analysis or low-SNR residuals.
It may also be possible for system-specific properties to impact the sodium absorption lines, such as stellar activity or variability, or changes in atmospheric structure. However, we cannot make a conclusive statement regarding this and further investigation is required to fully understand the implication of these results \citep[e.g.][]{slanger2005, oza2019, zak2019, gebek2020, borsa2021}.

\subsection{Winds}
\label{sec:winds}
In addition to the line depths, it is possible to measure deviations from the rest frame wavelength of the sodium doublet using the centroids of the Gaussian fits. Since efforts were made to shift the spectra into the planetary rest frame and minimise offsets due to systemic and stellar radial velocities, any deviation from the rest frame position can be interpreted as being due to winds in the planetary atmosphere. Many previous studies have identified and analysed blueshifted absorption lines corresponding to winds with velocities of a few km~s$^{-1}$ \citep[e.g.][]{snellen2010, louden2015, wyttenbach2015, casasayas-barris2019, seidel2020a}. General circulation models (GCMs) for hot Jupiters also predict that the hottest region of the atmosphere can be advected downwind from the substellar point, and atmospheric flow may be dominated by a super-rotating eastward equatorial jet \citep[e.g.][]{showman2009, showman2013, kempton2012, rauscher2014}.

The measured velocity offsets of the Gaussian centroids from the rest frame position of the sodium doublet are displayed in Table~\ref{tab:winds}. The bottom panel of Figure~\ref{fig:line_ratios_winds} shows these measurements for the weight-combined spectra, plotted against $\xi$. It is noted that many of these results have large uncertainties that overlap with zero within \textasciitilde~1$\sigma$, making it difficult to justify a clear detection of atmospheric winds for individual planets. However, when comparing these values across the sample, an important trend is that all planets in this survey (except WASP-79b) have a net blueshifted velocity of a few km~s$^{-1}$, consistent with predictions from GCMs \citep{koll2018}.
These results imply that there may be common and fundamental processes regarding winds and atmospheric dynamics in gas giants within a broad \mbox{\textasciitilde~1000--4000~K} temperature range \citep{heng2015b, showman2015, showman2020}. The uncertainty and inconclusiveness for WASP-79b is likely due to inaccuracies from only having one night of data and few out-of-transit frames, and further observations of this planet would be beneficial to help to improve this result and confirm the trend across the sample of planets.

\begingroup
\renewcommand{\arraystretch}{1.2} 
\begin{table}
    \centering
    \caption{Wind velocities for the ten gas giant planets analysed in this survey, measured from the offsets of the sodium D-lines from their rest frame position.}
    \label{tab:winds}
    \begin{tabular*}{\columnwidth}{l@{\extracolsep{\fill}}cc}
    \toprule
    \multirow{2}{*}{Planet} & Velocity (km s$^{-1}$)    & Velocity (km s$^{-1}$)    \\
                            & (Weight-combined)         & (Unweighted)              \\
    \midrule
    WASP-69b                & $-2.5\pm3.6$              & $-3.4\pm4.9$              \\
    HD~189733~b             & $-1.8\pm1.7$              & $-2.1\pm1.4$              \\
    WASP-21b                & $-2.6\pm2.4$              & $-1.2\pm2.4$              \\
    WASP-49b                & $-4.9\pm4.9$              & $-2.2\pm2.1$              \\
    WASP-79b                & $0.5\pm3.4$               & $-2.7\pm3.9$              \\
    WASP-76b                & $-1.5\pm1.7$              & $-9.0\pm2.6$              \\
    MASCARA-2b              & $-3.1\pm1.9$              & $-2.9\pm2.2$              \\
    WASP-121b               & $-2.9\pm3.4$              & $-11.0\pm4.8$             \\
    WASP-189b               & $-3.9\pm1.1$              & $-1.8\pm0.9$              \\
    KELT-9b                 & $-4.1\pm2.9$              & $-3.6\pm3.0$              \\
    \bottomrule
    \end{tabular*}
\end{table}
\endgroup

\section{Summary and Conclusion}
\label{sec:conclusions}

We conducted a homogeneous survey of sodium in a diverse sample of ten gas giants using high-resolution transmission spectroscopy from the HARPS and HARPS-N spectrographs. We outlined the key steps involved to remove telluric and interstellar contamination, correct for radial velocities, and remove artefacts due to Centre-to-Limb Variation and the Rossiter-McLaughlin effect. A consistent method was employed for all observations to prevent the results from being influenced by differences in assumptions and other factors between independent analyses.

We confirmed the presence of sodium in nine planets which have been studied in previous work, and reproduced relevant results for verification. We reported a new detection of Na in the atmosphere of WASP-79b, with line depths of $-1.12 \pm 0.23~\%$ (D2) and \mbox{$-0.85 \pm 0.22~\%$} (D1).
The fitted Gaussian profiles are a good approximation for the line depths, however further investigation using other atmospheric models may promote valuable scientific discussion \citep[e.g.][]{ehrenreich2006, wyttenbach2015, pino2018, oza2019, gebek2020, hoeijmakers2020, seidel2020a, seidel2021}.

The planetary transmission spectra were computed by combining the in-transit residual spectra in the rest frame of the planet, using both simple and weighted averages. For the second approach, weights of $1/\sigma^2$ were applied to each flux value which reduces the contribution of noisy pixels, particularly in regions where very little flux is measured due to deep stellar spectral lines. This is an important consideration to make, since low-SNR regions may be responsible for masking out or falsely enhancing the atmospheric features \citep{seidel2020b, seidel2020c}. We compared the effects of both combination methods on the measured sodium absorption features for each planet. In the majority of cases, either method was sufficient and the line depths were comparable within 1$\sigma$. However, there are certain scenarios (e.g. HD~189733~b, WASP-49b, and WASP-69b) where it is important to consider the quality of the data and the position of the planetary features within the residual spectra. First, if there are relatively few in-transit frames, the overall SNR of the combined transmission spectrum will be lower than for other targets, making it more difficult to measure the absorption lines. Next, the planetary features may be falsely enhanced or masked out if they are located inside the band of low-SNR stellar residuals -- this is not usually a problem for planets with higher radial velocities. Finally, further uncertainty may be added if there is significant overlap between the location of the planetary features and the spurious signals induced by CLV and the RM effect. Due to the inconsistent results in these certain situations, it is important to use the same method across all observations if there is a goal to compare the population of exoplanet atmospheres and look for trends. We therefore confirm that the weighted average approach that has been commonly employed in recent high-resolution transmission spectroscopy studies is most appropriate.

We performed a search for trends linking sodium absorption to the bulk properties of the ten gas giants in our sample. The measured absorption depths were expressed in terms of the relative height of the atmospheric sodium layer to the planetary radius, $h_{\text{Na}} = H_{\text{Na}}/R_{\text{p}}$. Only the D2 line depths were used for the trend analysis since they were clearly resolved and measured for each planet (a separate analysis of the D1 lines showed similar trends with larger uncertainties). 
We assessed the dependence of $h_{\text{Na}}$ on the equilibrium temperature and surface gravity (two properties directly associated with the incident irradiation and the atmospheric scale height) through the quantity \mbox{$\xi$ = $(T_{\text{eq}}/1000~\text{K})(g/g_{\text{J}})$}. 
An exponential trend of the form
\begin{equation}
    h_{\lambda} = a\text{e}^{-b\xi} + c ~,
    \label{eqn:height_trend_again}
\end{equation}
provided a good fit to the data, where $a=1.70 \pm 1.04$, $b=5.04 \pm 1.63$, and $c=0.113 \pm 0.013$. 
As shown in Figure~\ref{fig:height_ratio_trend}, the exponential trend is asymptotic to $h_{\text{Na}}\sim0.113$ for $\xi\gtrsim1.25$. Therefore, our results suggest that planets with $\xi\gtrsim1.25$ may have an upper limit on the relative height of the sodium layer of $0.113~R_{\text{p}}$. 
For planets in our sample with $\xi\lesssim1$, $h_{\text{Na}}$ increases exponentially with decreasing $\xi$. 

A notable exception to this trend is the planet WASP-79b for which the observed relative height is significantly higher than the trend. However, the measurement was based on only one available transit and the overall SNR of the combined transmission spectrum is lower than most of the other planets. WASP-79b is therefore a good candidate for follow-up observations, and it would be beneficial to combine more transits to improve the accuracy of the measurements. Additionally, further observation of WASP-69b and other low-$\xi$ planets would add valuable data to this work and help to constrain the parameters of the exponential fit.

We compared the empirical trend to theoretical expectations assuming that the observations probe the same number of scale heights in the atmosphere and that the terminator is isothermal with a temperature $T_{\text{eq}}$ across all the targets. Under these idealised conditions, the theoretical estimates of the normalised atmospheric heights follow a comparable trend to the observed values for some of the targets, but are significantly deviant for several others. This suggests that the idealised assumption is unlikely to be realistic across the diverse sample of planets considered in this work. Deviations from the trend may also inform us about the diversity of atmospheric processes among giant exoplanets. Our results therefore provide new impetus for further observations and theoretical investigations to assess the suggested trend and deviations thereof.

The proposed trend allows one to estimate which planets may or may not show significant sodium absorption within the limits of high-resolution observations from instruments such as HARPS. This could compliment other observational assessments for the purpose of selecting targets and optimally planning observations.
We performed a short case study on five additional planets with a range of $\xi$ values, equilibrium temperatures, and surface gravity. The low-$\xi$ planets \mbox{K2-232~b} ($\xi=0.36$), HAT-P-60~b ($\xi=0.40$), and \mbox{KELT-4~A~b} ($\xi=0.59$) had estimated D2 line depths of 0.69~\%, 0.44~\%, and 0.49~\% respectively. These depths are large enough to be verified with observations of two or three transits of each planet using HARPS-like facilities. In contrast, KELT-17b ($\xi=1.23$) and WASP-8b ($\xi=2.03$) had estimated D2 line depths of $0.21$~\% and $0.29$~\% respectively, which are comparable despite their distinct temperature and gravity. 
This highlights the importance of considering both properties simultaneously when trying to learn about the underlying trend. Due to the relatively shallow line depths and V-band apparent magnitudes greater than 9, sodium absorption is unlikely to be measurable for KELT-17b and WASP-8b using HARPS, which is supported by findings from recent work \citep{wyttenbach2017, stangret2021}.

We also investigated trends in the sodium D2/D1 line ratios. Seven planets had ratios consistent with 1, and three had ratios greater than 1. 
Due to the size of the uncertainties, we cannot make a conclusive statement about the line ratios using these results alone, and combining further observations would be beneficial.
Other studies have shown how line ratios can be used to learn about atmospheric structure \citep[e.g.][]{oza2019, gebek2020}, and our results may provide a useful basis for further research in this area. However, it is important to note that line ratios have been found to vary when observed with different instruments \citep[e.g. WASP-121b:][]{cabot2020, hoeijmakers2020, borsa2021}. 
Absorption lines could therefore be impacted by instrumental systematics, resolving power, low-SNR residuals, or weather effects. Other physical mechanisms may also contribute towards these variations, such as stellar activity or changes in atmospheric structure -- further investigation is required to fully understand these effects.

Several authors have previously identified absorption lines that are blueshifted with respect to their rest frame position, implying the presence of atmospheric winds \citep[e.g.][]{wyttenbach2015, louden2015, casasayas-barris2019}. For each planet in our sample, we measured the velocity offsets of the Gaussian centroids from the rest frame positions of the sodium lines. The velocities for several of the targets had large uncertainty ranges which overlapped with a net zero offset, making it difficult to justify the presence of winds for individual planets. However, when comparing across the sample, the general trend was that all offsets were blueshifted overall by a few km~s$^{-1}$. This consistency suggests the existence of strong day-night winds in a diverse range of exoplanets (as predicted by theoretical models), implying that there may be common underlying processes that govern the atmospheric dynamics of strongly irradiated gas giants \citep{rauscher2014, heng2015b, showman2020}.

The scientific community is gradually building a wealth of knowledge about the characteristics of exoplanet atmospheres. Observing time dedicated towards acquiring high-resolution spectra of transiting systems has facilitated the detection of numerous chemical species. This work has highlighted our ability to use measurements of a common absorption feature to search for trends among a population of irradiated gas giants, and understand how planetary properties can affect the extent and structure of their atmospheres. Applying the same methodology and combining our results with data from other current and future-generation high-resolution spectrographs such as ESPRESSO \citep{pepe2013}, EXPRES \citep{jurgenson2016}, HARPS-3 \citep{thompson2016}, and ANDES \citep{marconi2016}, will improve our evaluation of these trends and allow for more extensive comparison of different atmospheric species. This is a critical step towards enhancing our knowledge about the underlying processes of exoplanet atmospheres.

\section*{Acknowledgements}

We thank the anonymous referee for their thoughtful comments which helped to improve the quality of this manuscript.
AL acknowledges support from the Science and Technology Facilities Council (STFC), UK. AL thanks Simon Hodgkin and Annelies Mortier for insightful discussion about the HARPS spectrograph, instrumental effects, and data reduction.
This research has made use of the NASA Exoplanet Archive, which is operated by the California Institute of Technology, under contract with the National Aeronautics and Space Administration under the Exoplanet Exploration Program.

\section*{Data Availability}

All data used within this work are available through the ESO and TNG science archives. Observations with the HARPS spectrograph were obtained under programmes 072.C-0488(E), 079.C-0828(A), 079.C-0127(A), 087.C-0649(A), 096.C-0331(B), 090.C-0540(H), 0100.C-0750(A), 099.C-0898(A), 0101.C-0889(A), and 0103.C-0472(A). Observations with the HARPS-N spectrograph were obtained under programmes CAT16A\_130, CAT18A\_D1, CAT17A\_38, CAT18A\_34, CAT19A\_97, A35DDT4, and OPT18A\_38.


\bibliographystyle{mnras}
\bibliography{ms}


\appendix

\section{System Parameters}
\label{app:system_parameters}

\begin{landscape}
\begingroup
\renewcommand{\arraystretch}{1.5} 
\begin{table}
    \centering
    \caption{Stellar, planetary, and orbital parameters adopted in this work for HD~189733~b, KELT-9b, MASCARA-2b/KELT-20b, WASP-21b, and WASP-49b. 
        The quantities displayed are: V-band apparent magnitude ($m_{\text{V}}$), stellar mass ($M_\ast$), stellar radius ($R_\ast$), stellar RV semi-amplitude ($K_\ast$), effective temperature ($T_{\text{eff}}$), projected rotational velocity ($v\sin{i}$), surface gravity ($\log{g}$), metallicity ([Fe/H]), microturbulence ($v_{\text{mic}}$), macroturbulence ($v_{\text{mac}}$), planetary mass ($M_{\text{p}}$), planetary radius ($R_{\text{p}}$), planetary RV semi-amplitude ($K_{\text{p}}$), equilibrium temperature ($T_{\text{eq}}$), orbital inclination ($i_{\text{p}}$), sky projected obliquity ($\lambda$), period ($P$), mid-transit time ($T_\text{c}$), transit duration ($T_{14}$), semi-major axis ($a$), and systemic velocity ($v_{\text{sys}}$). 
        ${^\dagger}$~Derived using $K_{\text{p}} = -K_{\ast}(M_{\ast}/M_{\text{p}})$. 
        ${^\ddagger}$~Derived using $T_{\text{eq}} = T_{\text{eff}}\left[(1-f_\text{r})(1-A_\text{B})R_\ast^2/(2a^2)\right]^{1/4}$, assuming uniform heat redistribution ($f_\text{r} = 0.5$), zero bond albedo, and a circular orbit.
        $^*$~Due to the constraints on $K_\ast$ and $M_{\text{p}}$, this value is derived using the upper limit of both parameters. 
        $^\S$~Value is assumed. 
        References to listed parameters: 
        (A)~\citet{triaud2009}, 
        (B)~\citet{torres2008}, 
        (C)~\citet{boisse2009}, 
        (D)~\citet{stassun2017}, 
        (E)~\citet{bonomo2017}, 
        (F)~\citet{miller-ricci2008}, 
        (G)~\citet{moutou2020}, 
        (H)~\citet{agol2010}, 
        (I)~\citet{cegla2016}, 
        (J)~\citet{hoeijmakers2019}, 
        (K)~\citet{gaudi2017}, 
        (L)~\citet{borsa2019}, 
        (M)~\citet{talens2018}, 
        (N)~\citet{lund2017}, 
        (O)~\citet{casasayas-barris2019}, 
        (P)~\citet{ciceri2013}, 
        (Q)~\citet{bouchy2010}, 
        (R)~\citet{doyle2014}, 
        (S)~\citet{chen2020b}, 
        (T)~\citet{lendl2016}, 
        (U)~\citet{wyttenbach2017}, 
        (V)~\citet{lendl2012}.
        }
    \begin{tabular*}{1.34\textwidth}{l@{\extracolsep{\fill}}llllll}
    \toprule
    Parameter           & Unit                      & HD~189733~b                                           & KELT-9b                                           & MASCARA-2b/KELT-20b                                                                  & WASP-21b                                              & WASP-49b                                   \\
    \midrule
    \textbf{Star}       & & & & & \\
    $m_{\text{V}}$      & mag.                      & 7.67                                                  & 7.55                                              & 7.59                                                                                 & 11.590                                                & 11.352 \\
    $M_\ast$            & $M_{\sun}$                & 0.823 $^{+0.022}_{-0.029}$ $^{(\text{A})}$            & 1.978 $^{+0.023}_{-0.023}$ $^{(\text{J})}$        & 1.89 $^{+0.06}_{-0.05}$ $^{(\text{M})}$                                              & 0.890 $^{+0.079}_{-0.079}$ $^{(\text{P})}$            & 1.003 $^{+0.100}_{-0.100}$ $^{(\text{T})}$ \\
    $R_\ast$            & $R_{\sun}$                & 0.756 $^{+0.018}_{-0.018}$ $^{(\text{B})}$            & 2.178 $^{+0.011}_{-0.011}$ $^{(\text{J})}$        & 1.60 $^{+0.06}_{-0.06}$ $^{(\text{M})}$                                              & 1.136 $^{+0.051}_{-0.051}$ $^{(\text{P})}$            & 1.038 $^{+0.038}_{-0.036}$ $^{(\text{T})}$ \\
    $K_\ast$            & m s$^{-1}$                & 200.56 $^{+0.88}_{-0.88}$ $^{(\text{C})}$             & 276 $^{+79}_{-79}$ $^{(\text{K})}$                & $<322.51$ $^{(\text{N})}$                                                            & 37.2 $^{+1.1}_{-1.1}$ $^{(\text{Q})}$                 & 57.5 $^{+2.1}_{-2.1}$ $^{(\text{U})}$ \\
    $T_{\text{eff}}$    & K                         & 5052 $^{+16}_{-16}$ $^{(\text{D})}$                   & 9600 $^{+400}_{-400}$ $^{(\text{L})}$             & 8980 $^{+90}_{-130}$ $^{(\text{M})}$                                                 & 5800 $^{+100}_{-100}$ $^{(\text{Q})}$                 & 5600 $^{+150}_{-150}$ $^{(\text{D})}$ \\
    $v\sin{i}$          & km s$^{-1}$               & 3.5 $^{+1.0}_{-1.0}$ $^{(\text{E})}$                  & 111.40 $^{+1.27}_{-1.27}$ $^{(\text{K})}$         & 114 $^{+3}_{-3}$ $^{(\text{M})}$                                                     & 1.5 $^{+0.6}_{-0.6}$ $^{(\text{Q})}$                  & 0.9 $^{+0.3}_{-0.3}$ $^{(\text{V})}$ \\
    $\log{g}$           & $\log_{10}$(cm s$^{-2}$)  & 4.49 $^{+0.05}_{-0.05}$ $^{(\text{D})}$               & 4.093 $^{+0.014}_{-0.014}$ $^{(\text{K})}$        & 4.31 $^{+0.02}_{-0.02}$ $^{(\text{M})}$                                              & 4.277 $^{+0.026}_{-0.026}$ $^{(\text{P})}$            & 4.5 $^{+0.1}_{-0.1}$ $^{(\text{D})}$ \\
    $[$Fe/H$]$          & dex                       & $-0.03$ $^{+0.08}_{-0.08}$ $^{(\text{B})}$            & $-0.03$ $^{+0.20}_{-0.20}$ $^{(\text{K})}$        & $-0.02$ $^{+0.07}_{-0.07}$ $^{(\text{M})}$                                           & $-0.46$ $^{+0.11}_{-0.11}$ $^{(\text{P})}$            & $-0.23$ $^{+0.07}_{-0.07}$ $^{(\text{V})}$ \\
    $v_{\text{mic}}$    & km s$^{-1}$               & 1 $^{(\text{F})}$                                     & 1 $^{(\text{L})}$                                 & 2.0 $^{+0.5}_{-0.5}$ $^{(\text{M})}$                                                 & 1.2 $^{+0.1}_{-0.1}$ $^{(\text{R})}$                  & 0.9 $^{+0.2}_{-0.2}$ $^{(\text{V})}$ \\
    $v_{\text{mac}}$    & km s$^{-1}$               & 3.5 $^{(\text{G})}$                                   & 0 $^{\S}$                                         & 15.1 $^{+0.6}_{-0.6}$ $^{(\text{M})}$                                                & 3.57 $^{+0.73}_{-0.73}$ $^{(\text{R})}$               & 2.9 $^{+0.3}_{-0.3}$ $^{(\text{V})}$ \\
    \addlinespace
    \textbf{Planet}     & & & & & \\
    $M_{\text{p}}$      & $M_{\text{J}}$            & 1.138 $^{+0.022}_{-0.025}$ $^{(\text{A})}$            & 2.44 $^{+0.70}_{-0.70}$ $^{(\text{J})}$           & $<3.518$ $^{(\text{N})}$                                                             & 0.276 $^{+0.018}_{-0.018}$ $^{(\text{P})}$            & 0.399 $^{+0.030}_{-0.030}$ $^{(\text{U})}$ \\
    $R_{\text{p}}$      & $R_{\text{J}}$            & 1.138 $^{+0.027}_{-0.027}$ $^{(\text{B})}$            & 1.783 $^{+0.009}_{-0.009}$ $^{(\text{J})}$        & 1.83 $^{+0.07}_{-0.07}$ $^{(\text{M})}$                                              & 1.162 $^{+0.052}_{-0.052}$ $^{(\text{P})}$            & 1.198 $^{+ 0.047}_{- 0.047}$ $^{(\text{U})}$ \\
    $K_{\text{p}}$      & km s$^{-1}$               & $-152$ $^{+3}_{-3}$ $^{\dagger}$                      & $-234$ $^{+95}_{-95}$ $^{\dagger}$                & $-181$ $^{\dagger*}$                                                                 & $-186$ $^{+82}_{-82}$ $^{\dagger}$                    & $-151$ $^{+13}_{-13}$ $^{\dagger}$ \\
    $T_{\text{eq}}$     & K                         & 1200 $^{+15}_{-15}$ $^{\ddagger}$                     & 3670 $^{+150}_{-150}$ $^{\ddagger}$               & 2350 $^{+50}_{-50}$ $^{\ddagger}$                                                    & 1330 $^{+40}_{-40}$ $^{\ddagger}$                     & 1450 $^{+40}_{-40}$ $^{\ddagger}$ \\
    $i_{\text{p}}$      & deg                       & 85.710 $^{+0.024}_{-0.024}$ $^{(\text{H})}$           & 86.79 $^{+0.25}_{-0.25}$ $^{(\text{K})}$          & 86.15 $^{+28}_{-27}$ $^{(\text{N})}$                                                 & 88.28 $^{+0.90}_{-0.69}$ $^{(\text{S})}$              & 84.48 $^{+0.13}_{-0.13}$ $^{(\text{T})}$ \\
    $\lambda$           & deg                       & $-0.4$ $^{+0.2}_{-0.2}$ $^{(\text{I})}$               & $-85.78$ $^{+0.46}_{-0.46}$ $^{(\text{L})}$       & 0.6 $^{+4}_{-4}$ $^{(\text{M})}$                                                     & 8 $^{+26}_{-27}$ $^{(\text{S})}$                      & 54 $^{+79}_{-58}$ $^{(\text{U})}$ \\
    \addlinespace
    \textbf{System}     & & & & & \\
    $P$                 & days          & 2.21857567 $^{+0.00000015}_{-0.00000015}$ $^{(\text{H})}$  & 1.4811235 $^{+0.0000011}_{-0.0000011}$ $^{(\text{K})}$   & 3.47410196 $^{+0.00000106}_{-0.00000106}$ $^{(\text{O})}$                & 4.3225130 $^{+0.0000021}_{-0.0000021}$ $^{(\text{S})}$     & 2.7817362 $^{+0.0000014}_{-0.0000014}$ $^{(\text{T})}$ \\
    $T_\text{c}$        & BJD           & 2454279.436714 $^{+0.000015}_{-0.000015}$ $^{(\text{H})}$  & 2457095.68572 $^{+0.00014}_{-0.00014}$ $^{(\text{K})}$   & 2457503.120120 $^{+0.00018}_{-0.00018}$ $^{(\text{O})}$                  & 2454743.0419 $^{+0.0019}_{-0.0022}$ $^{(\text{Q})}$        & 2456267.68389 $^{+0.00013}_{-0.00013}$ $^{(\text{U})}$ \\
    $T_{14}$            & days          & 0.07527 $^{+0.00020}_{-0.00037}$ $^{(\text{A})}$      & 0.16316 $^{+0.00048}_{-0.00048}$ $^{(\text{K})}$  & 0.14882 $^{+0.00092}_{-0.00090}$ $^{(\text{N})}$                                     & 0.1426 $^{+0.0013}_{-0.0012}$ $^{(\text{S})}$         & 0.08918 $^{+0.00062}_{-0.00062}$ $^{(\text{T})}$ \\
    $a$                 & A.U.          & 0.03120 $^{+0.00027}_{-0.00037}$ $^{(\text{A})}$      & 0.03462 $^{+0.00110}_{-0.00093}$ $^{(\text{K})}$  & 0.0542 $^{+0.0014}_{-0.0021}$ $^{(\text{N})}$                                        & 0.0499 $^{+0.0013}_{-0.0013}$ $^{(\text{P})}$         & 0.03873 $^{+0.00130}_{-0.00130}$ $^{(\text{T})}$ \\
    $v_{\text{sys}}$    & km s$^{-1}$   & -2.2765 $^{+0.0017}_{-0.0017}$ $^{(\text{C})}$     & $-17.74$ $^{+0.11}_{-0.11}$ $^{(\text{J})}$       & $-21.3$ $^{+0.4}_{-0.4}$ $^{(\text{M})}$                                             & $-89.4499$ $^{+0.0013}_{-0.0013}$ $^{(\text{S})}$     & 41.7261 $^{+0.0011}_{-0.0011}$ $^{(\text{U})}$ \\
    \bottomrule
    \end{tabular*}
    \label{tab:parameters1-5}
\end{table}
\endgroup
\end{landscape}

\begin{landscape}
\begingroup
\renewcommand{\arraystretch}{1.5} 
\begin{table}
    \centering
    \caption{Stellar, planetary, and orbital parameters adopted in this work for WASP-69b, WASP-76b, WASP-79b, WASP-121b, and WASP-189b. 
        The quantities displayed are: V-band apparent magnitude ($m_{\text{V}}$), stellar mass ($M_\ast$), stellar radius ($R_\ast$), stellar RV semi-amplitude ($K_\ast$), effective temperature ($T_{\text{eff}}$), projected rotational velocity ($v\sin{i}$), surface gravity ($\log{g}$), metallicity ([Fe/H]), microturbulence ($v_{\text{mic}}$), macroturbulence ($v_{\text{mac}}$), planetary mass ($M_{\text{p}}$), planetary radius ($R_{\text{p}}$), planetary RV semi-amplitude ($K_{\text{p}}$), equilibrium temperature ($T_{\text{eq}}$), orbital inclination ($i_{\text{p}}$), sky projected obliquity ($\lambda$), period ($P$), mid-transit time ($T_\text{c}$), transit duration ($T_{14}$), semi-major axis ($a$), and systemic velocity ($v_{\text{sys}}$). 
        ${^\dagger}$~Derived using $K_{\text{p}} = -K_{\ast}(M_{\ast}/M_{\text{p}})$. 
        ${^\ddagger}$~Derived using $T_{\text{eq}} = T_{\text{eff}}\left[(1-f_\text{r})(1-A_\text{B})R_\ast^2/(2a^2)\right]^{1/4}$, assuming uniform heat redistribution ($f_\text{r} = 0.5$), zero bond albedo, and a circular orbit.
        $^\S$~Value is assumed. 
        References to listed parameters: 
        (A)~\citet{anderson2014}, 
        (B)~\citet{casasayas-barris2017}, 
        (C)~\citet{ehrenreich2020}, 
        (D)~\citet{west2016}, 
        (E)~\citet{seidel2019}, 
        (F)~\citet{brown2017}, 
        (G)~\citet{smalley2012}, 
        (H)~\citet{delrez2016}, 
        (I)~\citet{anderson2018}, 
        (J)~\citet{lendl2020}.
        }
    \begin{tabular*}{1.34\textwidth}{l@{\extracolsep{\fill}}llllll}
    \toprule
    Parameter           & Unit                      & WASP-69b                                          & WASP-76b                                          & WASP-79b                                                                      & WASP-121b                                              & WASP-189b                                 \\
    \midrule
    \textbf{Star}       & & & & & \\
    $m_{\text{V}}$      & mag.                      & 9.873                                             & 9.518                                             & 10.044                                                                        & 10.514                                                & 6.62 \\
    $M_\ast$            & $M_{\sun}$                & 0.826 $^{+0.029}_{-0.029}$ $^{(\text{A})}$        & 1.458 $^{+0.021}_{-0.021}$ $^{(\text{C})}$        & 1.39 $^{+0.06}_{-0.06}$ $^{(\text{F})}$                                       & 1.353 $^{+0.080}_{-0.079}$ $^{(\text{H})}$            & 1.887 $^{+0.057}_{-0.057}$ $^{(\text{I})}$ \\
    $R_\ast$            & $R_{\sun}$                & 0.813 $^{+0.028}_{-0.028}$ $^{(\text{A})}$        & 1.73 $^{+0.04}_{-0.04}$ $^{(\text{D})}$           & 1.51 $^{+0.04}_{-0.03}$ $^{(\text{F})}$                                       & 1.458 $^{+0.030}_{-0.030}$ $^{(\text{H})}$            & 2.36 $^{+0.03}_{-0.03}$ $^{(\text{J})}$ \\
    $K_\ast$            & m s$^{-1}$                & 38.1 $^{+2.4}_{-2.4}$ $^{(\text{A})}$             & 116.0 $^{+1.3}_{-1.4}$ $^{(\text{C})}$            & 90 $^{+8}_{-8}$ $^{(\text{F})}$                                               & 181 $^{+6.3}_{-6.4}$ $^{(\text{H})}$                  & 182 $^{+13}_{-13}$ $^{(\text{J})}$ \\
    $T_{\text{eff}}$    & K                         & 4715 $^{+50}_{-50}$ $^{(\text{A})}$               & 6329 $^{+65}_{-65}$ $^{(\text{C})}$               & 6600 $^{+100}_{-100}$ $^{(\text{G})}$                                         & 6459 $^{+140}_{-140}$ $^{(\text{H})}$                 & 7996 $^{+99}_{-99}$ $^{(\text{I})}$ \\
    $v\sin{i}$          & km s$^{-1}$               & 2.2 $^{+0.4}_{-0.4}$ $^{(\text{A})}$              & 1.48 $^{+0.28}_{-0.28}$ $^{(\text{C})}$           & 19.1 $^{+0.7}_{-0.7}$ $^{(\text{G})}$                                         & 13.5 $^{+0.7}_{-0.7}$ $^{(\text{H})}$                 & 93.1 $^{+1.7}_{-1.7}$ $^{(\text{J})}$ \\
    $\log{g}$           & $\log_{10}$(cm s$^{-2}$)  & 4.535 $^{+0.023}_{-0.023}$ $^{(\text{A})}$        & 4.128 $^{+0.015}_{-0.015}$ $^{(\text{D})}$        & 4.226 $^{+0.010}_{-0.010}$ $^{(\text{F})}$                                    & 4.242 $^{+0.011}_{-0.012}$ $^{(\text{H})}$            & 4.046 $^{+0.024}_{-0.024}$ $^{(\text{I})}$ \\
    $[$Fe/H$]$          & dex                       & 0.144 $^{+0.077}_{-0.077}$ $^{(\text{A})}$        & 0.23 $^{+0.10}_{-0.10}$ $^{(\text{D})}$           & 0.03 $^{+0.10}_{-0.10}$ $^{(\text{F})}$                                       & 0.13 $^{+0.09}_{-0.09}$ $^{(\text{H})}$               & 0.29 $^{+0.13}_{-0.13}$ $^{(\text{J})}$ \\
    $v_{\text{mic}}$    & km s$^{-1}$               & 0.7 $^{+0.02}_{-0.02}$ $^{(\text{A})}$            & 1.4 $^{+0.1}_{-0.1}$ $^{(\text{D})}$              & 1.3 $^{+0.1}_{-0.1}$ $^{(\text{G})}$                                          & 1.5 $^{+0.1}_{-0.1}$ $^{(\text{H})}$                  & 2.7 $^{+0.3}_{-0.3}$ $^{(\text{J})}$ \\
    $v_{\text{mac}}$    & km s$^{-1}$               & 0 $^{(\text{A})}$                                 & 4.0 $^{+0.3}_{-0.3}$ $^{(\text{D})}$              & 6.4 $^{+0.3}_{-0.3}$ $^{(\text{G})}$                                          & 6.6 $^{+0.6}_{-0.6}$ $^{(\text{H})}$                  & 0 $^{\S}$ \\
    \addlinespace
    \textbf{Planet}     & & & & & \\
    $M_{\text{p}}$      & $M_{\text{J}}$            & 0.260 $^{+0.017}_{-0.017}$ $^{(\text{A})}$        & 0.894 $^{+0.014}_{-0.013}$ $^{(\text{C})}$        & 0.85 $^{+0.08}_{-0.08}$ $^{(\text{F})}$                                       & 1.183 $^{+0.064}_{-0.062}$ $^{(\text{H})}$            & 1.99 $^{+0.16}_{-0.14}$ $^{(\text{J})}$ \\
    $R_{\text{p}}$      & $R_{\text{J}}$            & 1.057 $^{+0.047}_{-0.047}$ $^{(\text{A})}$        & 1.854 $^{+0.077}_{-0.076}$ $^{(\text{C})}$        & 1.53 $^{+0.04}_{-0.04}$ $^{(\text{F})}$                                       & 1.865 $^{+0.044}_{-0.044}$ $^{(\text{H})}$            & 1.619 $^{+0.021}_{-0.021}$ $^{(\text{J})}$ \\
    $K_{\text{p}}$      & km s$^{-1}$               & $-127$ $^{+12}_{-12}$ $^{\dagger}$                & $-198$ $^{+4}_{-4}$ $^{\dagger}$                  & $-154$ $^{+20}_{-20}$ $^{\dagger}$                                            & $-217$ $^{+14}_{-14}$ $^{\dagger}$                    & $-181$ $^{+19}_{-18}$ $^{\dagger}$ \\
    $T_{\text{eq}}$     & K                         & 960 $^{+20}_{-20}$ $^{\ddagger}$                  & 2210 $^{+30}_{-30}$ $^{\ddagger}$                 & 1720 $^{+30}_{-30}$ $^{\ddagger}$                                             & 2360 $^{+60}_{-60}$ $^{\ddagger}$                     & 2640 $^{+40}_{-40}$ $^{\ddagger}$ \\
    $i_{\text{p}}$      & deg                       & 86.71 $^{+0.20}_{-0.20}$ $^{(\text{A})}$          & 89.623 $^{+0.005}_{-0.034}$ $^{(\text{C})}$       & 86.1 $^{+0.2}_{-0.2}$ $^{(\text{F})}$                                         & 87.6 $^{+0.6}_{-0.6}$ $^{(\text{H})}$                 & 84.03 $^{+0.14}_{-0.14}$ $^{(\text{J})}$ \\
    $\lambda$           & deg                       & 0.4 $^{+2.0}_{-1.9}$ $^{(\text{B})}$              & 61.3 $^{+7.6}_{-5.1}$ $^{(\text{C})}$             & $-95.2$ $^{+0.9}_{-1.0}$ $^{(\text{F})}$                                      & $-257.8$ $^{+5.3}_{-5.5}$ $^{(\text{H})}$             & 86.4 $^{+2.9}_{-4.4}$ $^{(\text{J})}$ \\
    \addlinespace
    \textbf{System}     & & & & & \\
    $P$                 & days          & 3.8681382 $^{+0.0000017}_{-0.0000017}$ $^{(\text{A})}$    & 1.80988145 $^{+0.00000020}_{-0.00000028}$ $^{(\text{E})}$ & 3.662392 $^{+0.000004}_{-0.000004}$ $^{(\text{F})}$                   & 1.27492550 $^{+0.00000020}_{-0.00000025}$ $^{(\text{H})}$ & 2.7240330 $^{+0.0000042}_{-0.0000042}$ $^{(\text{I})}$ \\
    $T_\text{c}$        & BJD           & 2455748.83344 $^{+0.00018}_{-0.00018}$ $^{(\text{A})}$    & 2456107.85507 $^{+0.00034}_{-0.00034}$ $^{(\text{D})}$    & 2456215.4556 $^{+0.0005}_{-0.0005}$ $^{(\text{F})}$                   & 2456635.70832 $^{+0.00011}_{-0.00010}$ $^{(\text{H})}$    & 2458926.541696 $^{+0.000065}_{-0.000064}$ $^{(\text{J})}$ \\
    $T_{14}$            & days          & 0.0929 $^{+0.0012}_{-0.0012}$ $^{(\text{A})}$             & 0.1539 $^{+0.0008}_{-0.0008}$ $^{(\text{D})}$             & 0.1592 $^{+0.0008}_{-0.0008}$ $^{(\text{F})}$                             & 0.1203 $^{+0.0003}_{-0.0003}$ $^{(\text{H})}$         & 0.1813 $^{+0.0011}_{-0.0011}$ $^{(\text{I})}$ \\
    $a$                 & A.U.          & 0.04525 $^{+0.00053}_{-0.00053}$ $^{(\text{A})}$          & 0.0330 $^{+0.0002}_{-0.0002}$ $^{(\text{C})}$             & 0.0519 $^{+0.0008}_{-0.0008}$ $^{(\text{F})}$                             & 0.02544 $^{+0.00049}_{-0.00050}$ $^{(\text{H})}$      & 0.05053 $^{+0.00098}_{-0.00098}$ $^{(\text{J})}$ \\
    $v_{\text{sys}}$    & km s$^{-1}$   & $-9.62826$ $^{+0.00023}_{-0.00023}$ $^{(\text{A})}$       & $-1.073$3 $^{+0.0002}_{-0.0002}$ $^{(\text{D})}$          & 4.99 $^{+0.46}_{-0.06}$ $^{(\text{F})}$                                   & 38.350 $^{+0.021}_{-0.021}$ $^{(\text{H})}$           & $-24.452$ $^{+0.012}_{-0.012}$ $^{(\text{I})}$ \\
    \bottomrule
    \end{tabular*}
    \label{tab:parameters6-10}
\end{table}
\endgroup
\end{landscape}

\twocolumn
\section{RM and CLV corrections}
\label{app:rmclv}
\begin{figure}
    \centering
    \includegraphics[width=\columnwidth]{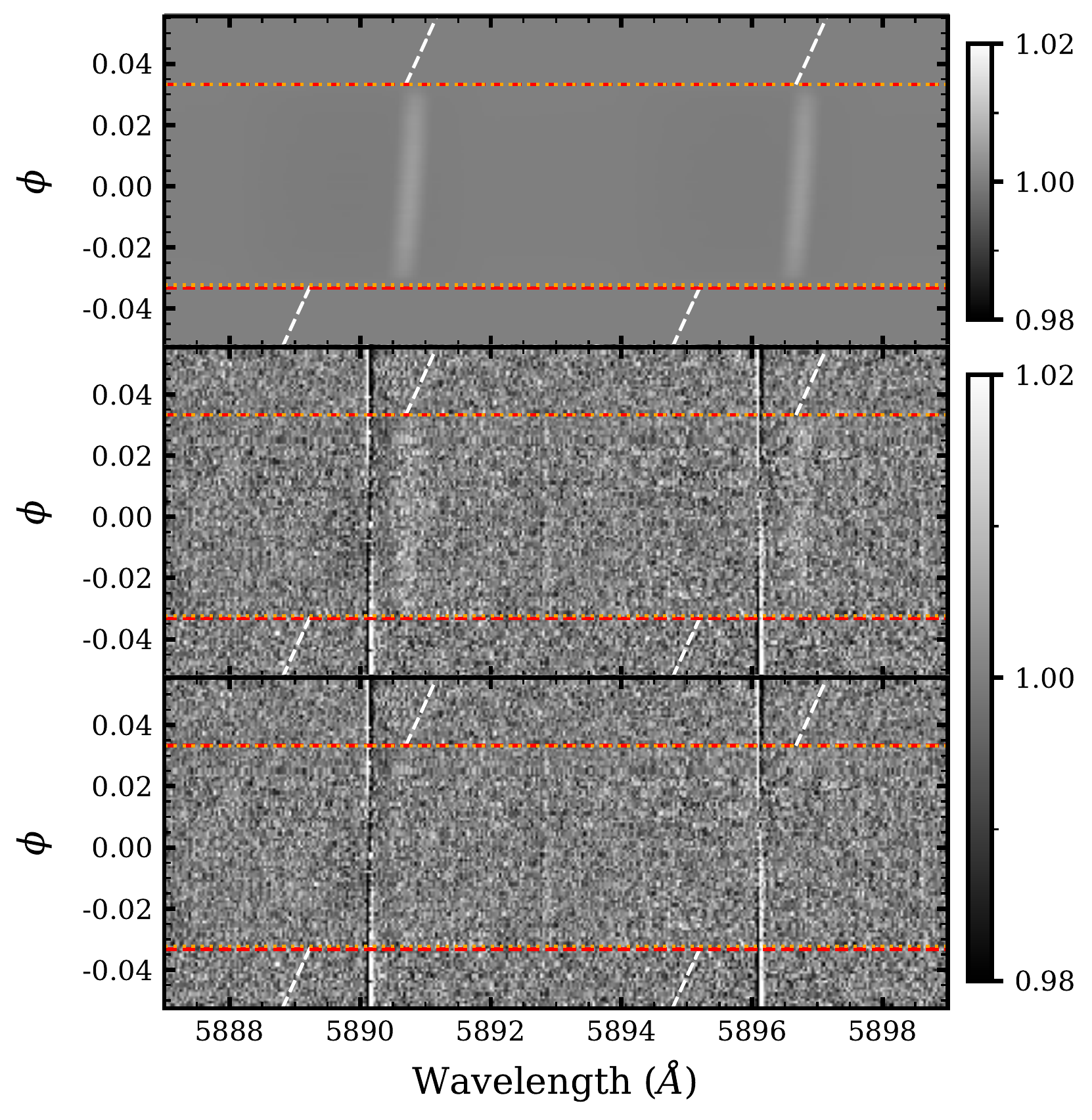}
    \caption{RM and CLV corrections for WASP-189b (night W189-N3). \textit{Top panel:} jointly modelled CLV and RM effects. \textit{Middle panel:} residual spectra which have not been corrected. By comparing this to the model, we can clearly see false artefacts within the data. \textit{Bottom panel:} result after dividing the residual spectra by the model. The false artefacts have been removed down to the noise level. We note that there are vertical bands of low/high residuals within the data. This is due to imperfect removal of ISM contamination, but it does not affect our overall results (see section~\ref{sec:method_tellurics} for further explanation).}
    \label{fig:rmclv_correction}
\end{figure}

\begin{figure}
    \centering
    \includegraphics[width=\columnwidth]{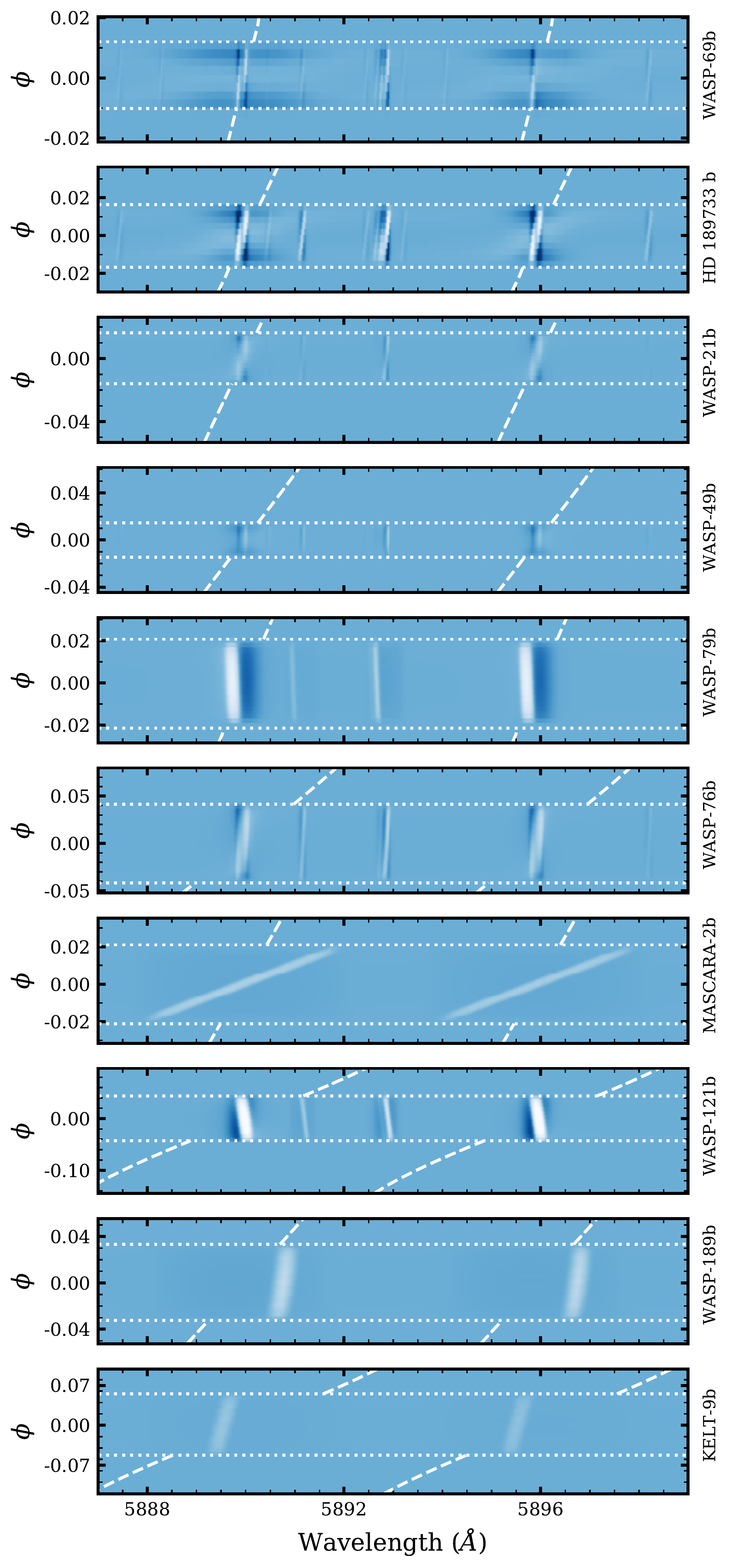}
    \caption{Models describing the combined CLV and RM effects for all ten planets (ordered by increasing $T_{\text{eq}}$), computed for all observed phases on one night. The minimum and maximum values (0.99 and 1.01) are the same in each panel. The horizontal dotted lines indicate the phases that are defined as "fully-in-transit", and the white dashed line highlights the trail of the atmospheric feature in the planetary rest frame. The models are very diverse, with each planet having unique RM and CLV features.}
    \label{fig:rmclv_models}
\end{figure}

Here we provide useful figures to describe how the spectra may be adversely affected by CLV and the RM effect. Figure~\ref{fig:rmclv_correction} shows how a model can successfully remove false artefacts from the data. Figure~\ref{fig:rmclv_models} shows the RM/CLV models for all ten planets analysed in this survey, using one night for each planet as an example.

\bsp	
\label{lastpage}
\end{document}